\documentclass[journal]{vgtc}                     

\onlineid{1507}

\vgtccategory{Research}

\title{Write, Rank, or Rate: Comparing Methods for \\ Studying Visualization Affordances}

\author{%
  \authororcid{Chase Stokes${}^\star\!\!$}{0000-0001-7644-9021},
  \authororcid{Kylie Lin${}^\star\!\!$}{0009-0002-2823-3992}, and 
  \authororcid{Cindy Xiong Bearfield}{0000-0002-1451-4083}
}

\authorfooter{
  \item[${}^\star\!\!$] Chase Stokes and Kylie Lin contributed equally to this work as co-first authors.
     Chase Stokes is with University of California, Berkeley.
  	E-mail: cstokes@ischool.berkeley.edu
  \item
  	Kylie Lin and Cindy Xiong Bearfield are with Georgia Institute of Technology.
        E-mail: klin368, cxiong@gatech.edu
}

\abstract{%
A growing body of work on visualization affordances highlights how specific design choices shape reader takeaways from information visualizations. However, mapping the relationship between design choices and reader conclusions often requires labor-intensive crowdsourced studies, generating large corpora of free-response text for analysis. To address this challenge, we explored alternative scalable research methodologies to assess chart affordances. We test four elicitation methods from human-subject studies: free response, visualization ranking, conclusion ranking, and salience rating, and compare their effectiveness in eliciting reader interpretations of line charts, dot plots, and heatmaps. Overall, we find that while no method fully replicates affordances observed in free-response conclusions, combinations of ranking and rating methods can serve as an effective proxy at a broad scale. The two ranking methodologies were influenced by participant bias towards certain chart types and the comparison of suggested conclusions. Rating conclusion salience could not capture the specific variations between chart types observed in the other methods. To supplement this work, we present a case study with GPT-4o, exploring the use of large language models (LLMs) to elicit human-like chart interpretations. This aligns with recent academic interest in leveraging LLMs as proxies for human participants to improve data collection and analysis efficiency. GPT-4o performed best as a human proxy for the salience rating methodology but suffered from severe constraints in other areas. Overall, the discrepancies in affordances we found between various elicitation methodologies, including GPT-4o, highlight the importance of intentionally selecting and combining methods and evaluating trade-offs.
}


\teaser{
  \centering
  \includegraphics[width=\linewidth, alt={A table depicting main findings.}]{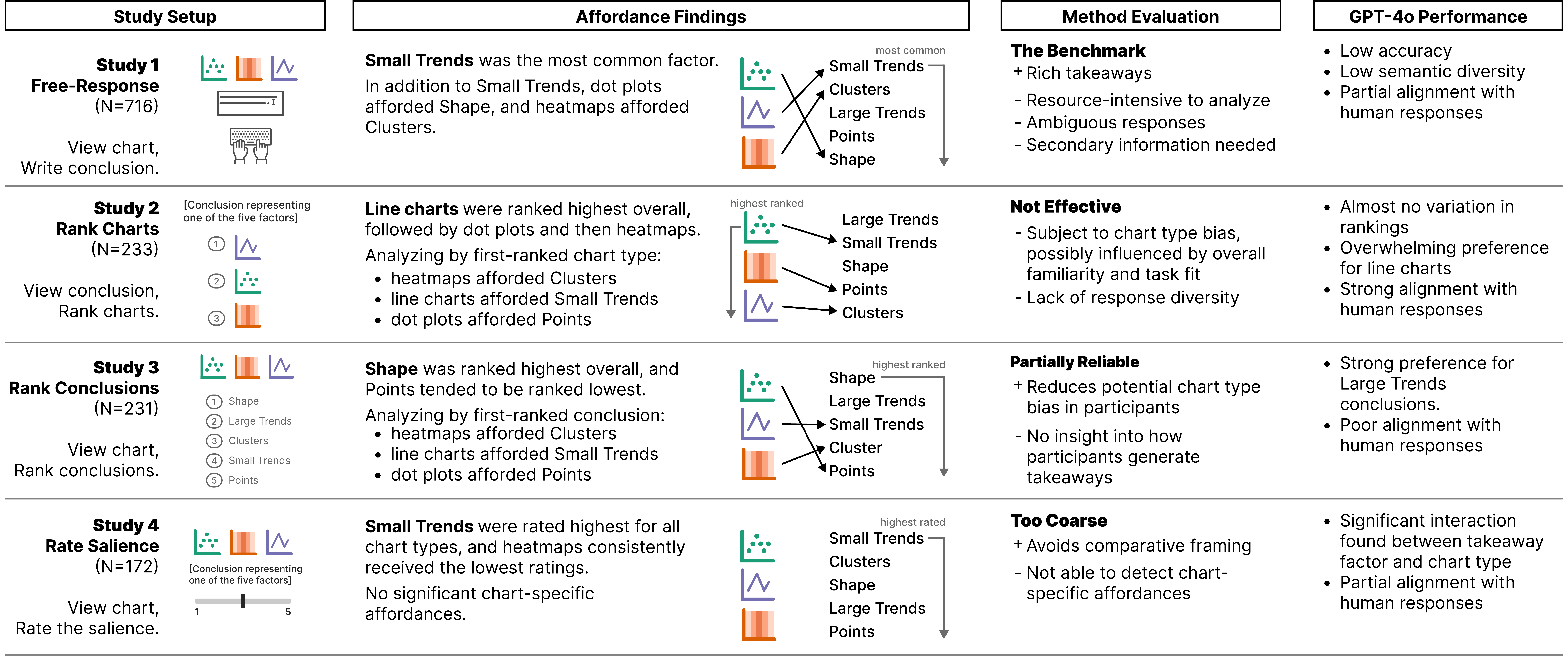}
  \caption{%
  	In this work, we compare four methods for eliciting visualization affordances from three canonical chart types (dot, line, and heatmap). In a case study, we also evaluate GPT-4o on its ability to capture affordances by comparing its output to human responses.%
  }
  \label{fig:teaser}
}

\graphicspath{{figs/}{figures/}{pictures/}{images/}{./}} 

\usepackage{tabu}                      
\usepackage{booktabs}                  
\usepackage{lipsum}                    
\usepackage{mwe}                       

\usepackage{mathptmx}   

\usepackage{enumitem}
\usepackage{fontawesome}
\usepackage{colortbl}
\usepackage{url}
\usepackage{float}
\usepackage{pifont}
\usepackage{graphicx}
\usepackage{multirow}
\usepackage{framed}
\PassOptionsToPackage{dvipsnames,svgnames,x11names}{xcolor}
\usepackage{xcolor}
\usepackage{tabularx}
\usepackage[normalem]{ulem}
\useunder{\uline}{\ul}{}

\newcommand{\pheading}[1]{\vspace{3px}\noindent\textbf{#1}}

\definecolor{salmon}{RGB}{232, 125, 114}
\definecolor{shadecolor}{RGB}{232,232,232}
\definecolor{humancolor}{RGB}{139, 0, 0}
\definecolor{gptcolor}{RGB}{255, 114, 86}
\arrayrulecolor{lightgray} 

\newcolumntype{R}[1]{>{\raggedright\arraybackslash}p{#1}}

\newcommand{\smallTrends}{\texttt{Small Trends}\xspace}
\newcommand{\largeTrends}{\texttt{Large Trends}\xspace}
\newcommand{\clusters}{\texttt{Clusters}\xspace}
\newcommand{\points}{\texttt{Points}\xspace}
\newcommand{\shape}{\texttt{Shape}\xspace}

\newenvironment{tightItemize}{\begin{itemize} \itemsep
-2.5pt}{\end{itemize}}

\usepackage{tabularx}
\usepackage{tabu}                      
\newcolumntype{P}[1]{>{\arraybackslash}p{#1}}

\usepackage{xurl}

\usepackage{booktabs}

\begin{document}


\firstsection{Introduction}

\maketitle

The `right' visualization design allows a viewer to derive clear takeaways from even complex data~\cite{gleicher2011visual}. 
Choices in visual encoding, such as spatial arrangements or color manipulations, can significantly influence what people compare and take away from data~\cite{bearfield2023does, fygenson2023arrangement}. 
For example, bar charts depicting income distributions naturally emphasize a comparison of bar heights.
Viewers will tend to focus on grouping and averaging incomes, often leading to an overly general takeaway: everyone in one group earns more than those in another~\cite{holder2022dispersion}. 
This can potentially reinforce stereotypes about group differences.
In contrast, a jittered dot plot of the same data shifts attention to individual data points and within-group variability, leading to more nuanced takeaways and reducing the likelihood of stereotypical judgments.

Visualization researchers have long sought to map the relationship between design choices and reader takeaways, including what data points they compare, patterns they notice, and decisions they make~\cite{boy2015suggested, fygenson2023arrangement}. 
We refer to these relationships as \textit{visualization affordances}~\cite{stokes2024delays, bertini2020shouldn}.
A visualization affordance is the unique link between a design choice and what readers take away from the presented information~\cite{crouser2012affordance}.

Understanding visualization affordances typically requires researchers to conduct extensive empirical studies and collect large corpora of qualitative data on human responses~\cite{fygenson2023arrangement}. 
Interpreting these responses is not only labor-intensive but also fraught with ambiguity. 
For example, lexically, a term like `spread' may refer to variability (i.e., how many clusters of points are dispersed across the chart) or range (i.e., the difference between the highest and lowest values). 
Semantically, a statement such as ``compared to Paper B, Paper A received a higher score from Reviewers 1 and 2,'' could imply either a combined comparison across papers or an individual comparison between scores from each reviewer~\cite{xiong2021visual}.
These ambiguities in human responses describing their takeaways require manual reviews to be deciphered, which limits the scalability of affordances studies, and subsequently, our systematic understanding of affordances in visualizations. 

Therefore, 
\textbf{we investigate alternative research methods to collect chart takeaways at scale, aiming to increase the efficiency of studying visualization affordances.}
We compare four takeaway elicitation methods: free responses, chart ranking, conclusion ranking, and salience rating. 
If the last three methods generate comparable outcomes of chart takeaways to the in-depth free response method, they offer a scalable and efficient way to study visualization affordances. 

As an additional exploration for efficient research data collection, we conduct a brief case study on how a Large Language Model (LLM) would respond to similar visualization interpretation prompts.
In recent years, researchers have explored using such models as proxies for human subjects in empirical studies~\cite{dillion2023can}, including strategies for generating synthetic research data~\cite{horton2023large, gilardi2023chatgpt, hamalainen2023evaluating, argyle2023out}.
While LLMs offer powerful computational capabilities, their effectiveness as human proxies -- particularly in visualization research -- remains debated \cite{harding2024ai, wang2024aligned, bendeck2024empirical, hong2025llms}.
To contribute to the debate, we evaluate the efficacy of a state-of-the-art LLM (OpenAI’s GPT-4o~\cite{achiam2023gpt}) as a human proxy for visualization affordance studies, 
across all variations of affordance elicitation methodologies examined for humans. 
We identify the limitations and capabilities of using LLMs as a research tool to study visualization affordances.

\vspace{1mm}
\noindent \textbf{Contribution:} We contribute: 
\textbf{(1)} A comparison of four research methodologies, 
exploring their trade-offs and implications for effectively studying visualization affordances.
\textbf{(2)} Five data-driven factors to categorize human takeaways from visualization: \points, \smallTrends, \shape, \largeTrends, and \clusters, based on results from a series of human-subject studies. 
\textbf{(3)} Suggested affordances for different chart types according to converging evidence across four methods. \textbf{(4)} A case study evaluating the capability of OpenAI’s GPT-4o to match the behavior of human participants in visualization affordance studies.

\section{Related Works}

Visualization design shapes the type of information people extract and the inferences they draw from data. Foundational work in exploratory data analysis by Tukey~\cite{tukey1977exploratory} and empirical studies by Cleveland and McGill~\cite{cleveland1984graphical} demonstrated that different graphical encodings vary in their effectiveness at conveying specific data patterns. 
Building on these early insights, visualization researchers have extensively examined how our visual system enables rapid extraction of aggregate statistics from visualizations~\cite{szafir2016four}, supporting tasks such as judging correlations in scatterplots~\cite{rensink2010perception, harrison2014ranking} and assessing probabilities~\cite{padilla2020powerful}.

\subsection{Visualization Affordances}
\label{affordanceBackground}

Recent work has defined visual affordances as the ``perceivable possibilities for visual tasks''~\cite{fygenson2023arrangement} that a visualization presents to a reader. 
Even basic design choices, such as selecting a chart type, can alter the affordances of a visualization and, therefore, what a viewer takes away from the data~\cite{cleveland1984graphical}. 
Thoughtful design choices can strengthen the communicative power of visualizations~\cite{padilla2018decision}, while poor design choices can obscure or distort the intended message in a visualization~\cite{burns2020evaluate, southwell2022defining}.  

Existing work has begun to synthesize common design best practices and empirical findings into structured guidelines for improving visual data communication~\cite{liu2014survey, munzner2014visualization}. 
For example, bar charts encourage readers to make magnitude comparisons (``A is larger than B''), while a line graph highlights changes over time, (``A is increasing at a higher rate than B'')~\cite{bearfield2023does}. 
Visualizations that aggregate data points (e.g., bar charts) can lead viewers to infer causality, whereas those that display probabilistic outcomes (e.g., scattplots), promote a better understanding of uncertainty~\cite{kay2016ish, holder2022dispersion}.
However, 
compared to more conventional visualizations, probabilistic visualizations (e.g., quantile dot plots) can undermine trust and confidence, likely due to unfamiliarity~\cite{yang2023swaying}.
In addition to chart types, color and shape selection also influence reader perceptions. 
Choosing colors that are most semantically aligned with viewers' mental models will increase information processing efficiency~\cite{setlur2015linguistic, schloss2024color}. 
Choosing the 'right' sets of shapes for categorical data simplifies comparative analyses in multi-class datasets~\cite{tseng2024shape, heider2011local}. 

Since visualization design influences patterns viewers see, it follows that these patterns influence viewer takeaways and decisions~\cite{zohrevandi2022design, yang2023swaying}.
In a study evaluating risk representations in a wildfire scenario, researchers found that participants were more likely to evacuate when using icon arrays with fewer icons compared to those with more icons~\cite{matzen2023numerical}. 
People seem to focus on the denominator of icon arrays, interpreting a larger number of icons as a `less risky' scenario~\cite{schapira2001frequency}. 

\subsection{Methods for Understanding Visualization Affordances}

In this work, we investigate methods for eliciting the information readers extract from a visualization, which we refer to as `chart takeaways.'
Battle and Ottley~\cite{battle2023insight} described chart takeaways as a type of insight, along with 
data facts (e.g., ``a unit of discovery''~\cite{saraiya2005insight}), hypotheses, or links connecting findings from data with existing knowledge (e.g., ``a complex, deep, qualitative, unexpected, and relevant assertion''~\cite{plaisant2008insight}). 

Strategies for studying chart takeaways typically include qualitatively coding textual responses~\cite{zacks1999bars} or visualizations drawn from textual descriptions~\cite{zacks1999bars}, as well as analyzing quantitative ratings of cognitive factors such as trust via Likert scales~\cite{south2022effective}. 
Coding text responses can be labor-intensive and contain lexical or semantic ambiguities~\cite{xiong2021visual}, while Likert scale ratings can fail to capture nuanced details about the specific takeaway messages. 
Most recently, Fygenson et al. have taken a chart-selection approach specifically for capturing visualization affordances, asking study participants to complete a presented message (using a fill-in-the-blank format) and then select one of four chart types that best represented the resulting message~\cite{fygenson2023arrangement}.

In our investigation, we begin by assuming that the best methodology for capturing visualization affordances is through a free-response task that provides rich information about ``the message(s) that readers tend to extract from a visualization''~\cite{fygenson2023arrangement}. 
To address the shortcomings of the free-response task as a benchmark, we evaluate three methodologies for capturing visualization affordances: ranking charts for a given conclusion, ranking conclusions for a given chart, and rating the saliency of a conclusion for a given chart. Each method is aimed towards being less time-consuming and subject to fewer ambiguities while still capturing visualization affordances.

\subsection{Large Language Models for Visualization Interpretation}
\label{sec:llm_rw}

Researchers have begun to explore the extent to which Large Language Models (LLMs), which are advanced statistical models pre-trained on vast corpora of natural language data, can enhance research workflows.
Recent advancements have paved the way to leverage LLMs for visual analytics~\cite{zha2023tablegpt, liu2023fill}, and visualization researchers have developed benchmarks for characterizing LLM performance across various tasks and evaluation criteria \cite{chen2024viseval}.
For instance, tools such as LEVA use LLMs to enhance analytics through three stages of visual analysis: onboarding, exploration, and summarization~\cite{zhao2024leva}. 

As another use case of LLMs, HCI and visualization researchers have proposed using LLMs as proxies for human participants in empirical studies~\cite{dillion2023can, harding2024ai}. 
While LLMs can approximate certain response patterns, such as binary ratings in moral judgments~\cite{dillion2023can}, behavior predictions in economic decision tasks~\cite{horton2023large, gilardi2023chatgpt}, and reactions to public health messages~\cite{das2025leveraging}, they often fall short in capturing the full nuance of human behavior. 
LLMs can also complete visualization literacy tasks~\cite{bendeck2024empirical, hong2025llms, xu2024exploring}, but they are prone to hallucinations and inconsistencies~\cite{goethals2024one, hamalainen2023evaluating}. 
They can sometimes struggle to accurately emulate human response to spatial manipulations and visual structures in visualizations~\cite{xiong2021visual}, and instead focus more on the dataset’s topic~\cite{wang2024aligned}.
Despite these limitations, LLMs offer promising opportunities to reduce study time and costs, while enhancing research efficiency, scalability, and applicability.

Given that a core motivation of this work is to explore tractable alternative methodologies for capturing visualization affordances while maintaining response quality comparable to free-response data, we devote a portion of our investigation to exploring how well a state-of-the-art LLM can perform in predicting visualization affordances.
In~\cref{sec:casestudy} we present a case study using GPT-4o~\cite{achiam2023gpt} out-of-the-box to determine a baseline of LLM capabilities for this context.

\section{Overview}

We compare methods for capturing the affordances of three canonical chart types~\cite{bertini2020shouldn}: dot plots, heatmaps, and line charts. Based on prior work,
we select the following four methods: free-response~\cite{zacks1999bars, xiong2021visual}, ranking chart types for a given conclusion~\cite{fygenson2023arrangement}, ranking conclusions for a given chart type, and rating the saliency of a given conclusion for a given chart~\cite{hearst2016evaluating, forsell2010heuristic}. 
\Cref{fig:teaser} shows a summary of the main findings.
The flow of this paper is as follows:

\textbf{Preliminary Study.} We first assess reader percepts across charts, characterizing types of affordances that readers derive from a visualization.
Results revealed five factors that broadly categorize readers' perceptions of visualizations, which we use in our subsequent experiments to characterize the types of visualization affordances. 

\textbf{Study 1: Free-Response.} We examine 
affordances through free-response to establish a benchmark mapping between chart types and takeaways. 
Participants reported takeaways 
using natural language, and we coded them using the factors identified in the preliminary study.

\textbf{Study 2: Rank Charts.} We examine affordances through a ranking task to compare against the benchmark established in Study 1. 
Participants ranked a set of charts based on how well each one highlighted a given takeaway. 
While some affordances aligned with those from Study 1, we also observed a moderate correlation between rankings and participants’ familiarity with the chart types.

\textbf{Study 3: Rank Takeaways.} Participants ranked five takeaways, each reflecting one of the five factors, based on a given chart.
The results were generally comparable to Study 1 with minor inconsistencies. 

\textbf{Study 4: Rate Salience.} Participants viewed chart-takeaway pairs and provided scalar ratings of salience. We did not observe any distinct chart affordances with this method.

\textbf{Case Study: GPT-4o.} Given recent exploration on whether large language models (LLMs) can serve as stand-ins for human participants in empirical studies~\cite{dillion2023can, harding2024ai, horton2023large, gilardi2023chatgpt, hamalainen2023evaluating, argyle2023out}, as well as the increasing incorporation of LLMs in visual analytics systems \cite{cui2024promises, zha2023tablegpt},
we investigate how a state-of-the-art LLM (GPT-4o) performs across all methodologies tested with human participants. 
We prompted GPT-4o with the same information provided to study participants.
Overall, human and GPT-4o responses diverged notably, echoing prior work on LLM limitations~\cite{wang2024aligned}, though some overlap suggests potential for improving their use as human proxies in visualization research.

\section{Preliminary Study: Chart Takeaway Factors}

We conducted a preliminary study to develop a framework for assessing visualization affordances. We used a crowdsourced set of conclusions for canonical chart types including dot plots, line charts, and heatmaps~\cite{bertin1983semiology}, since line charts and dot plots are common for time-series data, and heatmaps differ from position-based charts through the use of color encoding. 
We identified common patterns noticed in these visualizations through both theory-driven and data-driven qualitative coding. 
We grouped these patterns into five classes (factors) of takeaways based on a data-driven factor analysis.
These five factors defined our affordance space and served as the foundation for detecting visualization affordances throughout our investigations.

\subsection{Study Design}

\textbf{Participants.} We recruited 62 participants through the online crowdsourcing platform Prolific~\cite{palan2018prolific}, compensating them \$7.13 for a 45-minute survey. Participants completed the study in Qualtrics~\cite{snow2013qualtrics}.

\vspace{2pt}

\begin{figure*}[hbt!]
 \centering
 \includegraphics[width = \linewidth]{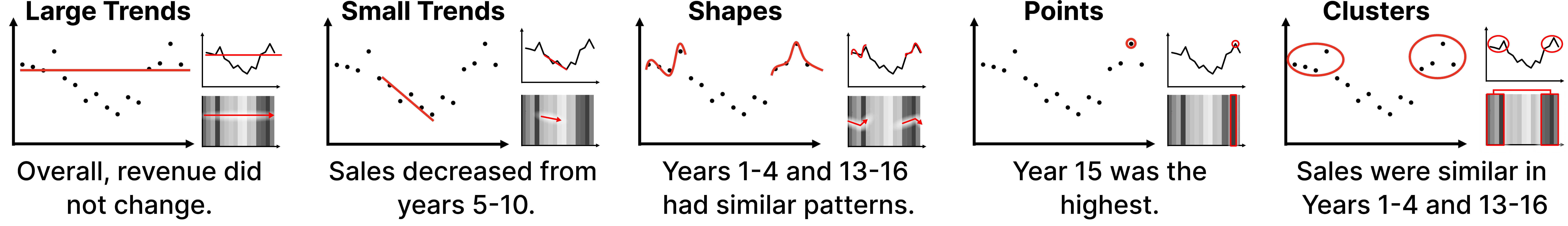}
 \caption{Types of conclusions surfaced from exploratory factor analysis along with an example conclusion and chart image highlighting the conclusion. 
 } 
 \label{fig:5factors}
\end{figure*}

\noindent \textbf{Stimuli.} We created stimuli for this study from three datasets, each with perceptually different trends, as recommended by Fygenson et al.~ \cite{fygenson2023arrangement}. 
The first dataset featured an increasing trend grouped into three groups of relatively stagnant revenue. The second dataset also had an increasing trend but was divided into six shorter groups of flat revenue (\Cref{fig:5factors} displays this dataset). The third dataset displayed sharp increases and decreases with no overall net change.
Charts depicted the revenue (y-axis) of fictional companies over 18 years (x-axis).

\vspace{2pt}

\noindent \textbf{Procedure.} First, participants completed two practice trials using unique datasets to familiarize themselves with the task. 
After completing the practice trials, each participant viewed six charts, evenly divided between dot plots, line charts, and heatmaps. 
We randomized the chart types and underlying datasets to control for order effects.
Participants never viewed two consecutive charts of the same type. 

For each chart, participants were asked to type their first takeaway.
After entering the takeaway, participants reported the range of years their responses referred to. 
We prompted them to repeat this process for a second and a third takeaway before moving on to the next chart.
Participants completed short distractor tasks in between the charts they viewed.
At the end of the study, participants reported demographic information. 
The survey also included three attention checks.
Participants who failed any attention checks were excluded from analysis.

\subsection{Category Schemes}

We categorized participant takeaways in two steps. 
We began with an open coding of participant responses, identifying 49 `percept codes' that referenced similar data features such as an increasing trend over time.
For example, the statements `Company sales have increased from the start' and `Revenue has doubled over 18 years' both describe upward trends and would therefore be assigned the same percept code.

We then applied a series of axial codes to the conclusions.
We repeated this with multiple coding schemes from the psychology literature and taxonomies in visualization research~\cite{brehmer2013multi, schulz2013design, shneiderman2003eyes, tory2004rethinking}, see below.
Specific codes and/or descriptions have been omitted for space considerations but can be found in supplementary materials.

\begin{tightItemize}
    \item \textbf{Unit Reason:} Constructed based on visual search processes \cite{treisman1982perceptual}; included information on the unit the participant selected (e.g., a data point, a subset of data), the property of the unit described (e.g., trend), and the operations performed with or across units (e.g., comparison of similar units). 
    \item \textbf{Perceptual Tasks:} Based on the \textit{perceptual task taxonomy} from Amar et al.~\cite{amar2005low}; in its original state, some tasks required participants to identify specific values 
    (e.g., filter, compute derived value). 
    We adapted this taxonomy to better encompass natural language conclusions: value, prediction, mean, extreme, range, distribution, anomaly, cluster, trend, difference, and compare.
    \item \textbf{Global vs. Local Perceptual Scope:} 
    Grounded in perceptual psychology research~\cite{navon1977forest, bearfield2023does}, which shows that viewers tend to first process broader global shapes before local components. This scheme categorized each conclusion as \textit{global} or \textit{local} in scope, 
    such as whether the response used the entire dataset, a subset of data, or a single point.
    \item \textbf{Intuitive Task:} Derived from a \textit{data-driven thematic analysis} that clustered into ten codes: 
    overall trend, pieces trend, shape, end point comparison, other point comparison, drastic change, within group value, within group relation, between group value, and between group relation. 
\end{tightItemize}

\subsection{Consolidating via Factor Analysis}
\label{sec:factor_analysis}

We collected 1,161 participant responses (390 from dot plots, 384 from heatmaps, and 387 from line charts). 
We conducted an exploratory factor analysis across the aforementioned coding schemes to consolidate the takeaways participants wrote as an expression of visualization affordance types.
The factor analysis was done in R Studio using the \texttt{Psych} R package~\cite{revelle2015package} and can be found in the supplementary materials.

We compared the empirical Bayesian information criterion (BIC) and model complexity of factor models consisting of 1 to 9 factors. 
We also examined how factors within each model correlated with each other and within each factor. 
Based on the balance of attributes, we decided to apply the five factor model to our codes. 
The five factors comprised (see~\cref{fig:5factors} for examples):

\begin{description}
    \item[\largeTrends:] Summarizes global trends across the full dataset, including predictions about future data points or overall averages.
    \vspace{-14px}
    \item[\smallTrends:] Highlights short-term changes or local fluctuations across subsets of the data, including large changes between adjacent points (e.g., spikes)
    \vspace{-5px}
    
    \item[\shape:] Description of overall shapes or patterns within the data distribution or subsets of data.
    \vspace{-5px}

    \item[\points:] Identifies individual data points and values, often the global or local maxima or minima.
    \vspace{-5px}
    \item[\clusters:] Groups data with similar values or visually similar regions within the chart.
\end{description}

\begin{figure*}[hbt!]
 \centering
 \includegraphics[width = 0.95\linewidth]{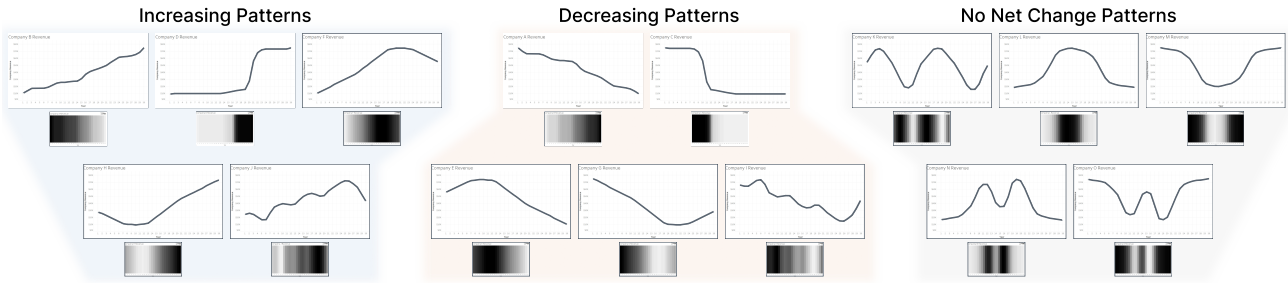}
 \caption{Stimuli datasets shown to participants. Each row displays a line chart and corresponding heatmap for the same data pattern. These 15 patterns were designed to span a range of trends and distributional shapes. Patterns fall into three groups, with five charts in each: increasing, decreasing, and no net change. All stimuli images can be found in the supplementary materials.
 }
 \label{fig:stimuli}
\end{figure*}

\subsection{Expanding Stimuli Set}
\label{sec:expanding_stimuli}

Informed by our preliminary study, we expanded the stimuli set to increase the generalizability of the following studies. 
We reviewed all bar, line, dot, and area charts from the MASSVIS ``targets393'' dataset~\cite{borkin2013makes}. 
The first author sorted them into groups based on similar data trends using a card sorting procedure, distilling 29 distinct data patterns. After further review, we collaboratively condensed this set to 15 patterns by removing overly similar variations.
These patterns include monotonic, oscillating, and step-like shapes to capture structural diversity.
We continued to use line charts, dot plots, and heatmaps in our follow-up studies to ensure comparability with prior work~\cite{bertini2020shouldn}.
We also abstracted away elements of MASSVIS designs such as text annotations, since text elements can influence reader interpretations \cite{stokes2022balance}.
Our final stimuli set consisted of 45 total charts (3 chart types x 15 datasets); examples can be seen in \cref{fig:stimuli}. 
This set allowed us to increase dataset variation while keeping the stimuli rooted in real-world data.

\section{Study 1: Free Response}

\begin{figure}[ht]
    \centering
 \includegraphics[width = 0.95\linewidth]{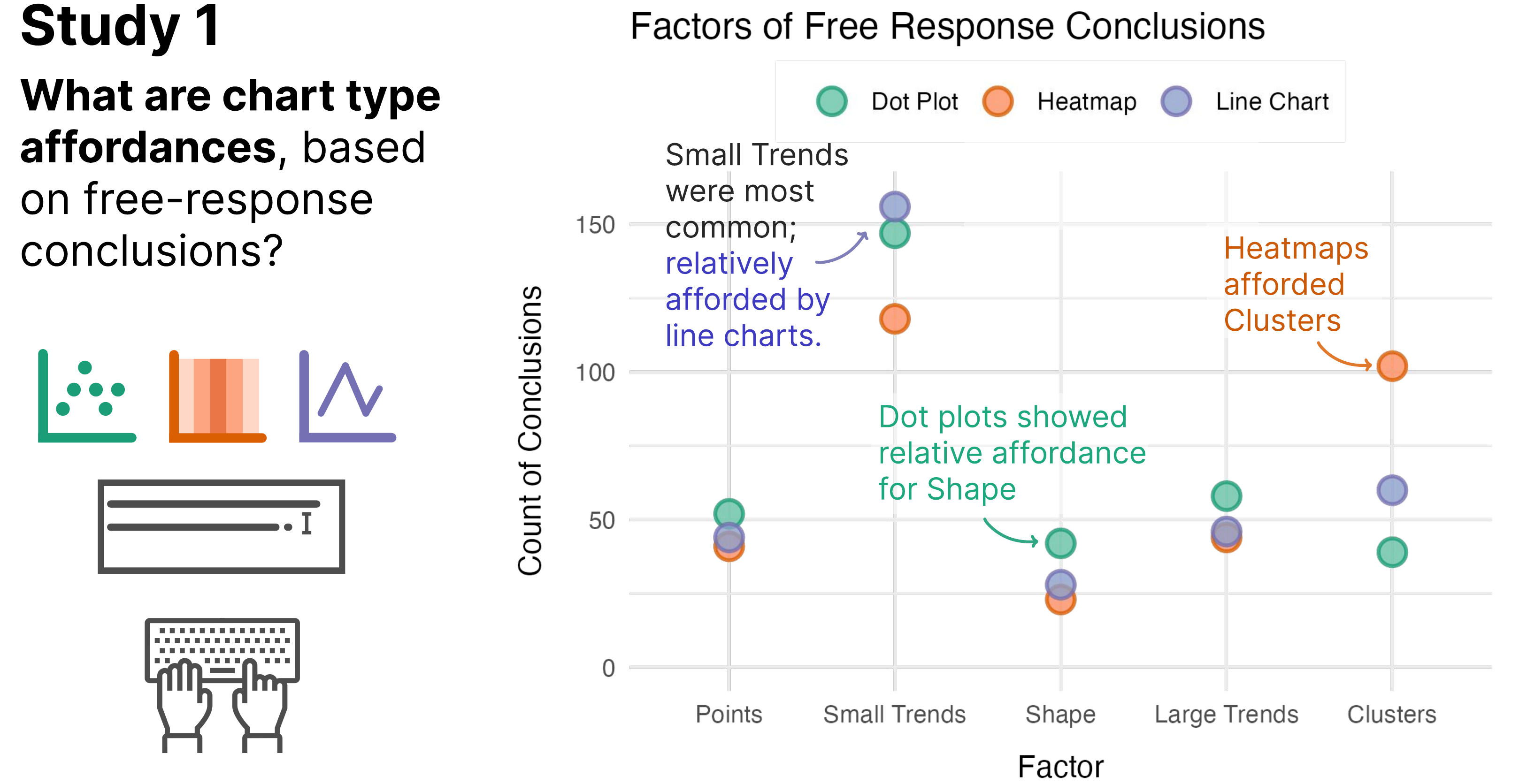}
    \caption{Study 1 results. \smallTrends were the most common, particularly for line charts. In addition to \smallTrends, dot plots afforded \shape, and heatmaps afforded \clusters.
    }
    \label{fig:study1_results}
\end{figure}

This study established a foundational benchmark for assessing visualization affordances by collecting free-response takeaways from participants. 
Participants were shown a randomly selected chart and responded with their primary takeaway. 
After coding and analyzing these responses, we identified distinct affordance patterns, shown in \cref{fig:study1_results}: heatmaps tended to elicit \clusters, dot plots emphasized \shape, and line charts afforded \smallTrends. 
These differences provide a critical baseline for comparing alternative elicitation methods.
Further details can be found in the supplementary materials.

\subsection{Participants and Procedure}

Using G*Power~\cite{faul2007g} for power analysis with pilot data, we determined that a sample size of $700$ participants would provide $85\%$ power at $\alpha = 0.05$.  
We recruited $770$ participants via Prolific~\cite{palan2018prolific}, filtering for native English speakers with an approval rate above $98\%$. 
After excluding people who failed an attention check or provided low-quality responses, we were left with $716$ participants. 
The majority ($57\%$) were between the ages of $25$ and $44$, and $41\%$ held a four-year degree.

Participants completed a Qualtrics survey, beginning with informed consent, followed by an attention check and a practice trial designed to familiarize them with writing natural language takeaways from charts.
Each participant then viewed one randomly selected chart from the set of 45 described in \cref{sec:expanding_stimuli} and reported their primary takeaway. 
Participants also provided the year(s) they focused on and the overall unit of focus (e.g., point, subset). 
Next, they answered demographic questions. 
Participants also reported their familiarity rating for each chart type, using a five-point scale from \textit{`1-Not familiar at all'} to \textit{`5-Extremely familiar'}. 
The survey took approximately four minutes, and participants were compensated \$0.80 on average.

To analyze responses, we coded participant takeaways according to the five factors identified in the preliminary study. Since real-world chart interpretation often includes errors, we included incorrect takeaways (e.g., incorrect values) in our analysis but focused on the takeaway factors to surface visualization affordances. To ensure reliability, $60\%$ of responses were double-coded by the authors ($\kappa=0.73$). After independently coding, the authors met to resolve discrepancies. 
If chart types afford different factors, we would expect to see that some charts consistently elicited certain factors more frequently than others. 
Identifying such patterns could emphasize that certain charts may naturally guide users toward specific takeaways.

\begin{figure*}[t!]
    \centering
 \includegraphics[width = 0.95\linewidth]{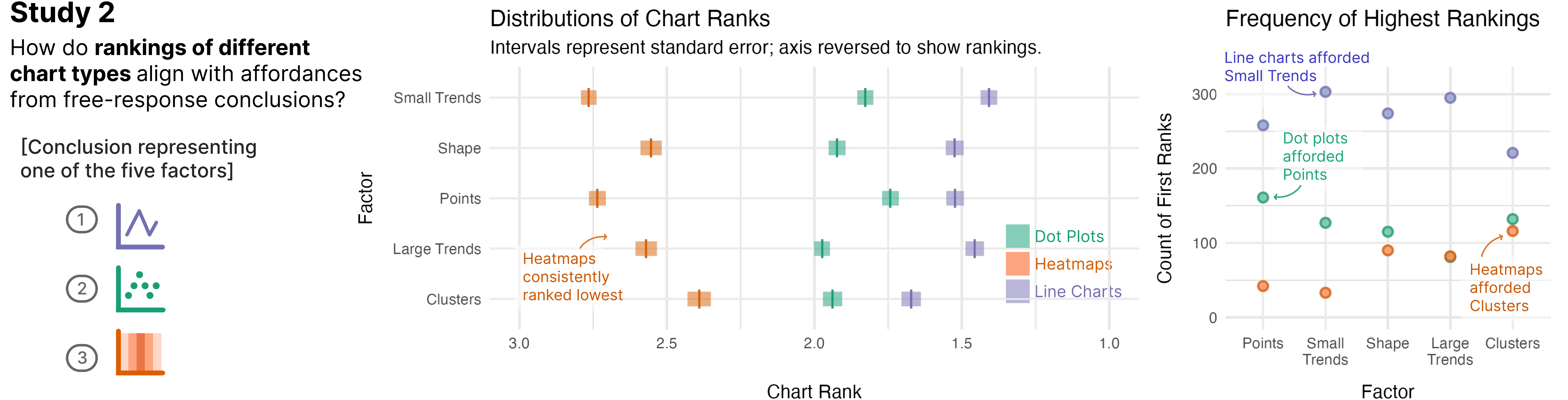}
    \caption{Study 2 results. Left: Distributions of chart type rankings; Line charts were ranked highest overall, followed by dot plots, then heatmaps. Right: We conducted additional analyses on the \textit{highest} ranked chart types, demonstrating affordances for heatmaps (\clusters) and line charts (\smallTrends) that aligned with Study 1. Other affordances (i.e., \points for dot plots) were in contrast to Study 1.
    }
    \label{fig:study2_results}
\end{figure*}

\subsection{Results}

Over a third of responses (34\%) contained multiple factors, with each takeaway receiving 1.4 factor codes on average. Only 6\% of the responses contained numerical values, suggesting that participants tended to describe overall patterns rather than specific data points.

\textbf{Factors by Chart Type.} Frequencies of the different factors for each chart type can be seen in \cref{fig:study1_results}.
Across all chart types, \smallTrends were the most frequent factor, followed by \clusters.
Through a Chi-squared analysis, we found significant variation in how different chart types shaped participant takeaways ($\chi^2 = 46.3, df = 8, p<0.001$).
Based on the standardized residuals, we extract specific differences in how each chart type afforded interpretations. 

Heatmaps elicited significantly more \clusters takeaways than other chart types ($R = 6.06$), suggesting that color encodings may highlight groupings of similar values. 
Dot plots were more likely to yield \shape takeaways ($R = 2.43$); participants may focus on spatial arrangements or patterns over time when using this chart type. However, \shape takeaways were not commonly generated by participants overall, including for dot plots. 
Line charts predominantly elicited \smallTrends takeaways ($R = 2.09$), aligning with their common use for tracking changes over time. The use of angle encodings may further enhance the salience of these changes.
\clusters takeaways appeared most with no-change charts; \smallTrends takeaways were frequent for decreasing trends and \largeTrends for increasing ones.
Overall, takeaways for line charts and dot plots were similar. These results support the notion that chart types afford different types of takeaways.

\subsection{Method Evaluation}

The free-response method provided rich data and mirrored how people naturally form takeaways in real-world contexts. 
This method uncovered distinct affordance patterns, validating the importance of chart selection in shaping user takeaways. 
However, analyzing these responses was resource-intensive, requiring extensive coding and analysis, including double-coding to ensure rigor and consistency. 

Additionally, some participant responses were ambiguous, leading the research team to seek clarification through collected secondary information (i.e., year ranges, unit of focus). 
On some occasions, this secondary information was necessary to determine the appropriate factor.
For example, if a response read ``Revenue peaked in Year 5,'' the reader could be focusing on the \textit{point} at Year 5 (\points) or the \textit{subset} of years that make up the peak (i.e., Years 4-6; \smallTrends).
Thus, we observed that supplementing free-response answers with additional clarification questions can facilitate more precise interpretations of text, though this does not make free-response a scalable approach.

\section{Study 2: Rank Charts}

This study examines the extent to which the method of ranking of different chart types can generate findings that align with affordances identified in Study 1. 
Participants ranked charts based on how well they conveyed a given message corresponding to one of five factors from the preliminary study. 
Details on the data, analysis, and participants are provided in the supplementary materials.

This method offers a structured, comparative way to assess affordances, similar to prior work \cite{fygenson2023arrangement}, where participants selected the visualization that \textit{most} clearly conveyed a given message.
We assessed whether certain chart types were consistently ranked higher for specific factors, which would suggest distinct affordances. 
Results indicated that chart rankings demonstrated a bias towards line charts and were correlated with chart type familiarity. Line charts were ranked highest overall, followed by dot plots, with heatmaps consistently ranked lowest. This suggests that while ranking tasks provide a different perspective on chart affordances, they may be influenced by participant familiarity with charts, among other features.

\subsection{Participants and Procedure}

A power analysis suggested that a sample size of $200$ would provide $85\%$ power at $\alpha = 0.05$. 
We recruited $270$ participants from Prolific~\cite{palan2018prolific}, applying exclusion criteria as in Study 1. 
The final sample consisted of $233$ participants with similar demographics to those in Study 1. 

Based on the five factors from the preliminary study (\cref{sec:factor_analysis}, we created five representative takeaways for each of the 15 data patterns described in~\cref{sec:expanding_stimuli} and shown in \cref{fig:stimuli}. 
These takeaways were designed to reflect those observed in Study 1. 
Examples include ``Company revenue decreased from Year 15 to Year 23'' (\smallTrends) and ``Every few years, revenue drops more and more'' (\shape). 

Participants 
were introduced to the chart types and ranking task via a Qualtrics survey. 
They were told there was no correct answer and to rank based on their subjective opinion.
After two practice trials, they were randomly assigned 10 sample takeaways. 
For each, they viewed the three chart types displaying the same dataset and ranked them in order of ``how well the charts highlighted the given message.'' 
Finally, participants reported demographics and chart type familiarity. 

This methodology was chosen as a more structured and quicker alternative to free-response.
However, unlike independent evaluations, ranking exposes participants to multiple chart options at once, which may shape judgments. 
While prior work focused only on the display that makes the message \textit{most} obvious \cite{fygenson2023arrangement}, we collected full rankings to explore broader interpretation patterns.
We analyze both overall ranking distributions and top-ranked charts.

\subsection{Results}

\textbf{Ranking Distributions.} 
Despite expectations that chart rankings would reveal affordance-driven differences, results showed that \textit{chart familiarity} was also associated with chart rankings. 
As illustrated in~\cref{fig:study2_results} line charts were ranked first on average across all factors (57.9\% of rankings).
This suggests that participants felt line charts afforded \textit{all} takeaways more strongly than other chart types.
Dot plots were typically ranked second (59.0\%), and heatmaps were ranked third (75.8\%). 
Using Durbin tests and post-hoc Conover testing with Holm correction~\cite{aickin1996adjusting}, these differences were significant across all factors ($p < 0.004$). 

Correlation analysis between familiarity ratings and chart rankings showed a moderate relationship ($\rho = -0.61, p<0.01$), suggesting that more familiar chart types tended to be ranked more highly.
Participants rated line charts as the most familiar ($Mean = 4.7, SD = 0.62$) and dot plots ($Mean = 4.2, SD = 0.95$), and heatmaps the least familiar ($Mean = 1.8, SD = 0.96$). 
While familiarity may offer a partial explanation for these rankings, it is only one contributing factor that we identified and does not indicate a causal relationship.
Line charts may genuinely be better suited to communicate temporal data, particularly when participants are prompted to choose the clearest option.

\textbf{Analysis of First-Ranked Charts.} We also conducted an analysis examining only the charts that participants ranked first, mirroring methods from Fygensen et al.~\cite{fygenson2023arrangement}. 
Given the strong effect of line charts observed in the overall ranking task, this analysis provides another window into participant responses. 
We find a significant difference in relative frequencies between chart type and takeaway types ($\chi^2 = 108.8$, $df = 8$, $p < 0.001$). 
An examination of standardized residuals highlights several notable patterns. Heatmaps were ranked first more often than expected for \clusters ($R = 6.12$), consistent with Study 1 findings.
Line charts were more frequently ranked first for \largeTrends ($R = 3.11$) and \smallTrends ($R = 3.63$), the latter also aligning with Study 1.  
Dot plots tended to be ranked first more than other chart types for \points ($R = 4.61$), suggesting they afford fine-grained comparisons of specific values. 

However, this pattern diverges from Study 1, where \shape was associated with dot plots. 
These findings suggest that while top rankings may reflect some underlying affordances, 
they do not consistently align with free-response conclusions.

\subsection{Method Evaluation}
Overall, we found that ranking tasks may be more effective for evaluating localized design variations (e.g., bar chart arrangements~\cite{fygenson2023arrangement}) than broader design choices like chart type. 
Although the first-ranked chart analysis provides clearer affordance signals than full-ranking data, line charts still accounted for $60\%$ of top selections, limiting the ability to draw general conclusions.
These preferences likely reflect a combination of factors, including familiarity and task fit, rather than affordance alone. If researchers aim to capture impressions of chart familiarity or prior experience, ranking tasks may provide useful signals. 
Overall, when analyzing ranking responses, it's important to recognize that rankings may reflect influences beyond affordances.


\section{Study 3: Rank Takeaways}

This study reverses the ranking charts method in Study 2. 
We examine the extent to which ranking takeaways based on how well they are represented in a given chart aligns with affordances identified with the free-response methods in Study 1. 
This method aimed to control for the chart type bias observed in Study 2 and provide a more clear evaluation of ranking as a methodology for assessing visualization affordances.
If chart types have distinct affordances, we would expect to see participants ranking the takeaways differently 
for each chart type. 
For example, to align with the affordances from Study 1, takeaways regarding \clusters would be ranked highly for heatmaps.

Our results partially support this outcome. \shape conclusions were ranked highly across all chart types, in contrast to Study 1 findings.
Relative rankings showed some affordance patterns across chart types but were less comprehensive than free-response insights from Study 1.

\subsection{Participants and Procedure}

Participant recruitment and power analysis followed the same procedures as Study 2. We recruited $270$ participants from Prolific, applying the same exclusion criteria.
The final sample consisted of $231$ participants with similar demographics to previous studies.

Participants completed a Qualtrics survey nearly identical to Study 2, except that they ranked \textit{takeaways} based on the five factors in the preliminary study, rather than chart types.  
In each trial, participants viewed one chart and ranked five takeaways, one per factor. 

This methodology was chosen to address the line chart bias observed in Study 2. By evaluating conclusions for only one chart type at a time, participants could not default to a preferred visualization. 
Instead, this task encouraged them to assess how well each chart conveyed different types of takeaways on its own.
We also conducted an additional analysis on the first-ranked conclusion for each chart.

\begin{figure*}[t!]
    \centering
 \includegraphics[width = 0.94\linewidth]{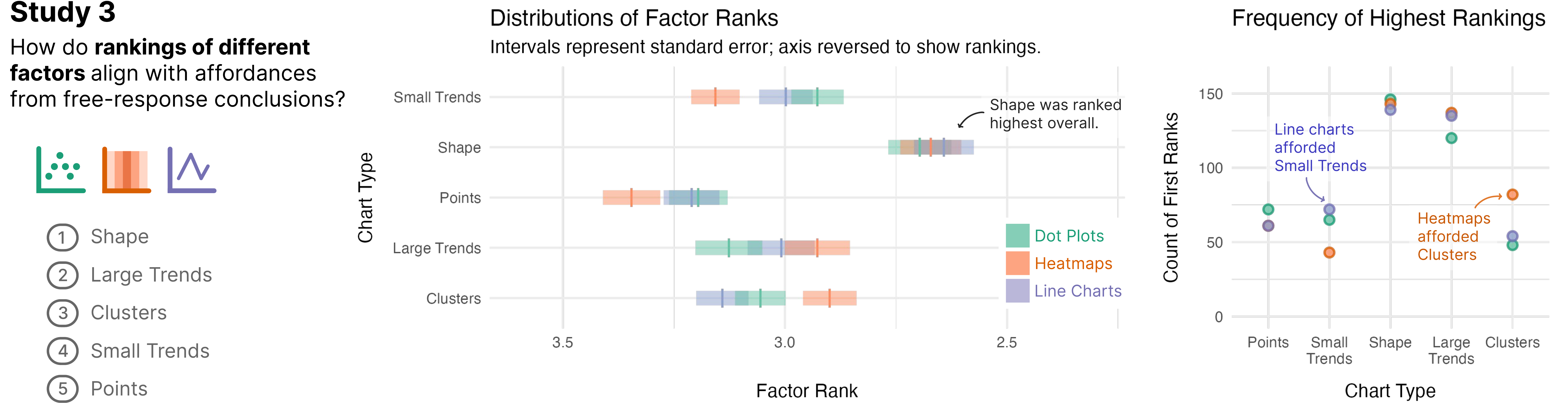}
    \caption{Study 3 results. 
    Left: Distributions of factor rankings. \shape was ranked highest overall, and many factors overlapped. \points tended to be ranked lowest. Right: Additional analyses on the \textit{highest} ranked factors, demonstrating affordances for heatmaps (\clusters) and line charts (\smallTrends) that aligned with Studies 1 and 2. Other affordances (i.e., \points for dot plots) were in contrast to Study 1 but aligned with Study 2.
    }
    \label{fig:study3_results}
\end{figure*}

\subsection{Results}

\shape conclusions were ranked as the most salient across all chart types, and \points conclusions were consistently the least salient.
This pattern diverges from the free-response findings in Study 1, where \shape conclusions were the least common and \smallTrends were the most common.
However, some ranking differences align with Study 1 findings, particularly for heatmaps and line charts.

\textbf{Ranking Distributions.} Rankings, shown in \cref{fig:study3_results}, varied significantly across chart types based on Durbin tests and post-hoc Conover testing with Holm correction.
For line charts, \shape and \smallTrends were ranked as the most salient. \shape was ranked significantly higher than all factors ($p < 0.013$) \textit{except} \smallTrends, which did not differ significantly from \shape ($p=0.075$) or any other factor (\points: $p=0.231$, \largeTrends: $p=1.00$, \clusters: $p=1.00$). 
This suggests that while \smallTrends were a salient feature of line charts, \shape appeared to be the dominant affordance.

When viewing heatmap visualizations, participants ranked \clusters significantly higher than \points ($p=0.009$), a finding not observed in line charts ($p = 1.00$) or dot plots ($p = 1.00$). 
This suggests that heatmaps may uniquely afford \clusters compared to other chart types, in line with the findings from Study 1. 
\largeTrends were also ranked higher than \points. This trend was also unique to heatmaps (line charts: $p=0.723$, dot plots: $p=1.00$), indicating some relative affordance for \largeTrends.
\shape was still ranked highest for heatmaps overall.

For dot plots, \shape was again the most salient feature. While this technically overlaps with findings from Study 1, the ubiquitous salience of \shape across all chart types indicates this was not a unique affordance of dot plots for this method.
The only significant pairwise comparison across the factors was between \shape and \points conclusions ($p=0.01$); there were no unique affordances for dot plots. 

\textbf{Analysis of First-Ranked Conclusions.} We again conducted a separate analysis on the first-ranked conclusions for each chart type, finding that some chart types were more likely to be associated with specific factors ($\chi^2 = 20.76$, $df = 8$, $p = 0.008$).
Standardized residuals reveal several modest but interpretable patterns. \clusters were again strongly associated with heatmaps ($R = 3.31$), and \smallTrends were relatively afforded for line charts compared to other chart types ($R = 2.00$). 
Both findings are in line with Study 1 and the first-ranked chart analysis from Study 2.
Dot plots showed only a slight association with \points ($R = 1.40$), consistent with Study 2 but not Study 1.
Pattern-related findings did not align with Study 1. \shape was most commonly ranked first for chart with no net change, \largeTrends were most afforded for decreasing trends, and \points were most common for increasing trends. 

\subsection{Method Evaluation}

Results from analyzing the full set of rankings from participants show closer alignment with Study 1 than Study 2 did, suggesting that \textbf{ranking takeaways within a single chart type} helped reduce the bias towards line charts.
Overall, these results suggest that the takeaway-ranking tasks offer a partially reliable window into chart-specific affordances, 
though results for dot plots remain inconsistent across methods. 

A limitation emerged when comparing results to Study 1: \shape was least common in free responses but most salient in rankings.
Ranking tasks may encourage participants to focus on differences between provided options, affording different takeaways than methods based on unconstrained interpretation. 
It may be that it is more cognitively complex to generate \shape takeaways, but that \shape tends to be the more salient takeaway when placed in comparison to other factors.


\section{Study 4: Rate Salience}
\label{sec:study4}

This study evaluated whether participants' ratings of takeaway salience aligned with the affordances observed in free-response interpretations from Study 1. 
Visual salience of important information is a common heuristic for evaluating visualization design~\cite{hearst2016evaluating, forsell2010heuristic, cabouat2024previs}, motivating our use of salience as a proxy for affordance. 
Rather than comparing the rankings of chart types or takeaways, participants viewed a single chart-takeaway pair and rated how visually salient the takeaway appeared, isolating the judgments of visual emphasis.

Our use of scalar salience ratings was informed by similar rating scales in visualization research; prior work used similar scales to measure trust \cite{pandey2023you, elhamdadi2023vistrust}, aesthetics \cite{he2022beauvis, ajani2021declutter}, decision confidence \cite{stokes2024delays, slomska2021different}, and display suitability for specific tasks \cite{gutierrez2019benefits}. 
If chart types afford different takeaways, specific takeaways would receive higher salience ratings than others for a given chart type. 
For example, \clusters would be rated as more salient for heatmaps
compared to other factors. 

\subsection{Participants and Procedure}

A power analysis using G*Power~\cite{faul2007g} indicated that a sample size of $171$ would provide $90\%$ power at $\alpha = 0.05$. 
We recruited $200$ participants from Prolific and applied the same filtering criteria,
landing with a final sample size of $172$ participants. 
The demographic profile was consistent with earlier studies.

Participants completed a Qualtrics survey that followed the same general format as the previous studies but was adapted for a salience rating task. 
Each trial presented a single chart and a caption describing a specific chart takeaway. ``Salient'' takeaways were defined as ones that `'stand out from other information in the chart,'' and participants were asked to rate how visually salient the takeaway appeared on a 5-point scale ranging from \textit{`Not at all salient'} (1) to \textit{`Very salient'} (5).

\begin{figure}[ht]
    \centering
 \includegraphics[width = 0.95\linewidth]{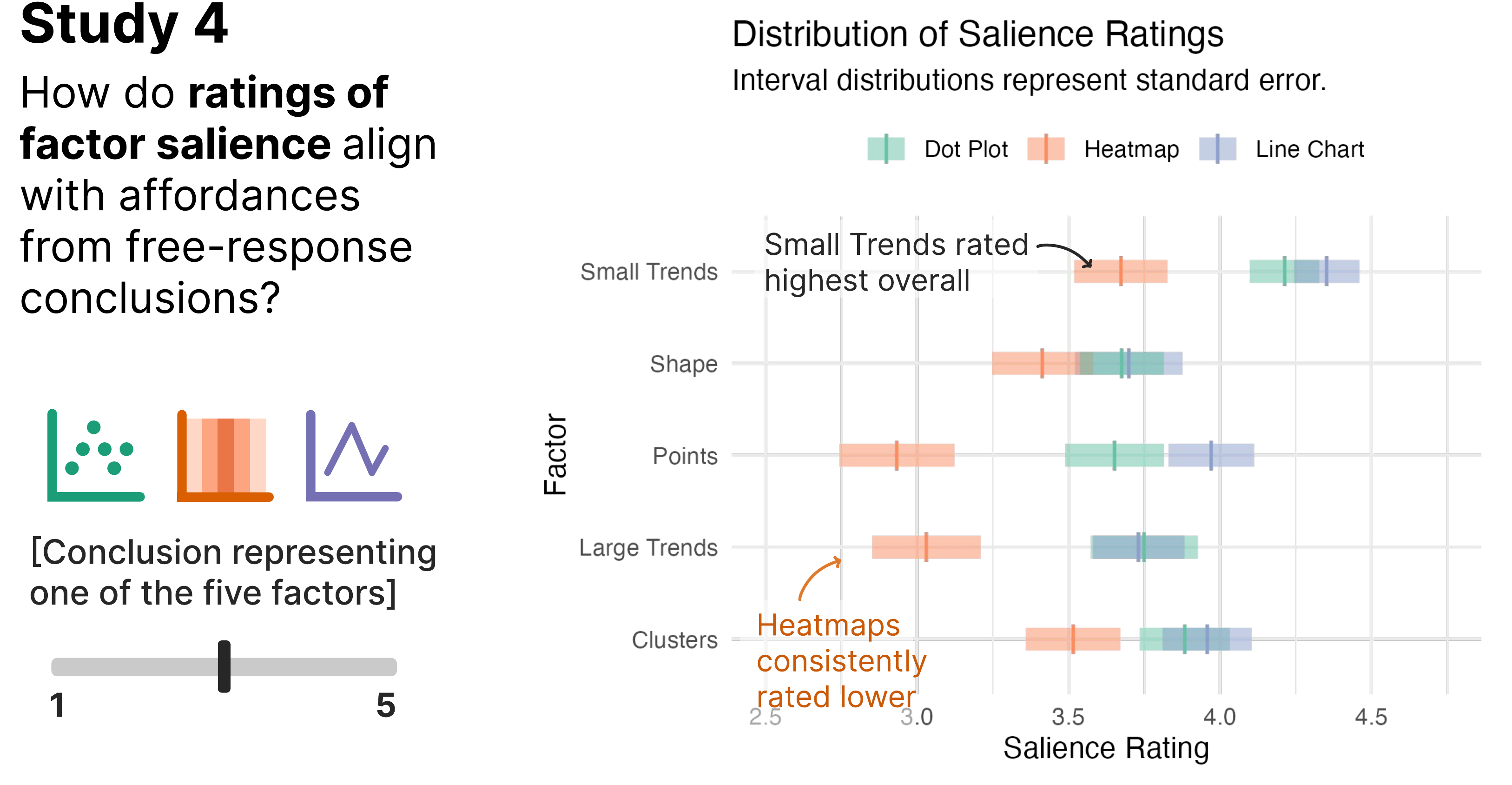}
    \caption{Study 4 results. There were no significant chart-specific affordances.
    \smallTrends were rated highest for all chart types, and heatmaps consistently received lower ratings than other chart types.}
    \label{fig:study4_results}
\end{figure}

\subsection{Results}

We analyzed salience ratings across chart-takeaway pairs
using an Analysis of Variance (ANOVA) test 
with post-hoc Tukey HSD testing. 
The ratings can be seen in~\cref{fig:study4_results}. 

\textbf{Ratings by Takeaways.} There was a significant main effect of takeaway types ($p<0.001$).
\smallTrends received the highest average salience ratings, significantly more than \largeTrends ($p < 0.001$), \points ($p < 0.001$), and \shape ($p = 0.001$). 
This finding is consistent with Study 1, where \smallTrends were the most common takeaway.

\begin{figure*}[hbt!]
    \centering
 \includegraphics[width = \linewidth]{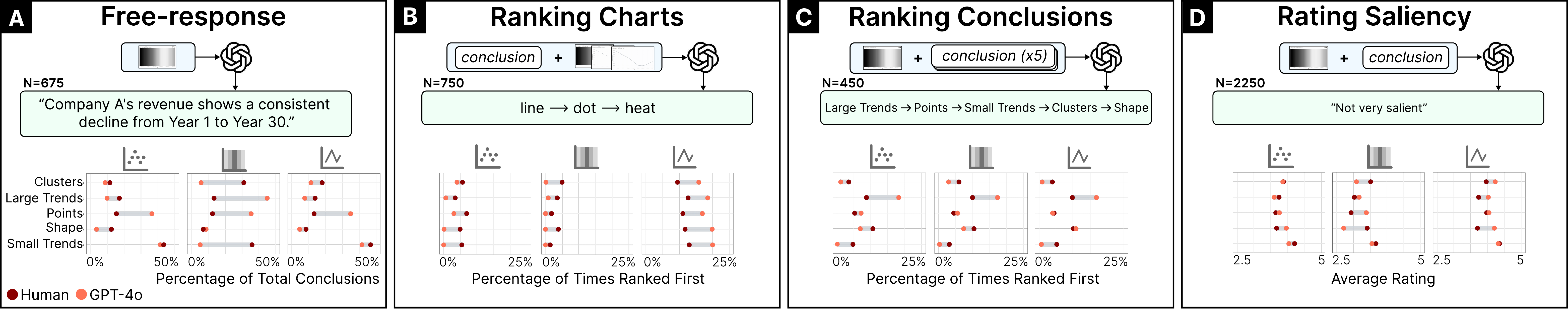}
    \caption{Overview of the approach and results of our case study with GPT-4o. 
    }
    \label{fig:casestudyoverview}
\end{figure*}

\textbf{Ratings by Chart Type.}There was also a significant main effect of chart type ($p<0.001$), with heatmaps receiving lower average saliency ratings across all factors in comparison to line charts ($p<0.001$) and dot plots ($p<0.001$). 
Charts with no net change also received lower average saliency ratings than both increasing and decreasing trends.
This mirrors the familiarity findings observed in Study 2, where heatmaps were consistently ranked lowest.
However, the interaction between chart type and takeaways was not significant ($p = 0.448$), indicating that 
participants did not perceive certain takeaways as more salient in specific chart types than others. These results fail to replicate the chart-specific affordances observed in Study 1.

Exploratory analysis showed that in heatmaps, the salience ratings for \shape and \clusters had overlapping standard errors with \smallTrends, suggesting comparable visual emphasis among these factors. 
For other charts, the salience ratings of the four non-dominant 
takeaways had overlapping standard errors, 
indicating no meaningful differences in perceived salience among them.

\subsection{Method Evaluation}

While this saliency rating method avoided the comparative framing of ranking tasks, the lack of interaction effects suggests it is not able to detect affordance differences present when participants generated their own takeaways in Study 1. 
The overall trends mirror some of the frequency patterns from Study 1, but the absence of chart-specific effects suggests this method may capture general salience patterns of datasets more than visualization-specific affordances.


\section{Case Study: Elicitation Methods with GPT-4o}
\label{sec:casestudy}

Thus far, we have explored three methods for capturing visualization affordances that are less time-consuming and more tractable than gathering and analyzing natural language responses. 
In light of recent academic interest in using LLMs as proxies for human research participants (see \cref{sec:llm_rw}), we next evaluate the ability of GPT-4o, a state-of-the-art LLM~\cite{achiam2023gpt}, to align with human responses when given prompts that closely match each elicitation method. 

\subsection{Approach}

We prompted GPT-4o with instructions that closely matched those given to our human participants in Studies 1-4, using parameters from recent human-GPT comparison studies in visualization research by Wang et al.~\cite{wang2024aligned}. We then conducted the same analysis used for human studies
and compared the results to human responses from Studies 1-4 (see~\cref{fig:casestudyoverview}).  Prompts can be found in supplementary materials.
This approach provides an exploration into LLMs as proxies for human subject participants.
Rather than applying extensive prompt engineering or comparing across models, our goal was to assess how well a straightforward, out-of-the-box approach performs. 
In doing so, we offer a novel and practical starting point for future visualization research on how to evaluate LLM output \cite{lin2025proxy}.
\subsection{Results}

Overall, we found that GPT-4o output most closely aligned with human study responses when prompted via the salience rating method (Study 4, \cref{sec:study4}). The model suffered from severe constraints for other methods. 
Free-response conclusions were largely inaccurate and lacking in semantic diversity, chart rankings overall contained extremely low variation, and affordances derived from conclusion rankings aligned poorly with human responses. 

\subsubsection{GPT-4o: Free-Response}
\label{sec:gptfreeresponse}

GPT-4o generated conclusions with many more inaccuracies compared to human responses. About 97\% of human conclusions were accurate, as opposed to 63\% of GPT-4o conclusions contained. We document types of inaccuracies from both humans and GPT-4o in~\cref{tab:inaccuracytable}, where multiple types of inaccuracies could apply to a single conclusion. 

\begin{table}[h]
\centering
\small 
\begin{tabularx}{\linewidth}{R{0.215\linewidth} lccl }
\toprule
\textbf{Inaccuracy Type} & \textbf{Source} & \textbf{Count} & \textbf{Percentage} & \textbf{By Chart Type}\\
\midrule

Inaccurate Trend(s) & \cellcolor{humancolor!40}Human & 9 & 1.2\% & 5 heat, 4 line \\
 & \cellcolor{gptcolor!40}GPT-4o & 83 & 12.3\% & 83 heat \\
 
Inaccurate Value(s) & \cellcolor{humancolor!40}Human & 7 & 0.9\% & 2 dot, 1 heat, 4 line\\ 
 & \cellcolor{gptcolor!40}GPT-4o & 137 & 20.3\% & 57 dot, 46 heat, 34 line\\ 

Likely Typo(s) & \cellcolor{humancolor!40}Human & 8 & 1.1\% & 5 dot, 1 heat, 2 line \\
& \cellcolor{gptcolor!40}GPT-4o & -- & -- & -- \\

Incorrect Cycle(s) & \cellcolor{humancolor!40}Human & -- & -- & -- \\
& \cellcolor{gptcolor!40}GPT-4o & 58 & 8.5\% & 29 dot, 1 heat, 28 line \\
 
\bottomrule
\end{tabularx}
\caption{Types of inaccuracies from human and GPT-4o free response.
}
\label{tab:inaccuracytable}
\vspace{-1.5em}
\end{table}

Beyond these inaccuracies, GPT-4o generated conclusions that varied significantly in structure and content from human responses. Responses generated by GPT-4o more often contained multiple takeaway factors (GPT-4o: $78\%$, Human: $34\%$).
GPT-4o was also more inclined to include specific data values in its conclusions  (GPT-4o: $26\%$, Human: $6\%$).
As a result of the increased prevalence of specific values, the \points factor was most common in conclusions from GPT-4o, as opposed to \smallTrends for humans. 

Affordances derived from GPT-4o free responses aligned only partially with affordances from human responses. We observed significant variations in factors across chart types ($\chi^2 = 46.3, df = 8, p<0.001$), with differences visualized in~\cref{fig:casestudyoverview}A. 
Across all chart types, GPT-4o generated more \points conclusions than humans. For dot plots in particular, GPT-4o did not capture \largeTrends or \shape with the same frequency as humans, although it generated a comparable proportion of \clusters. For heatmaps, GPT-4o aligned poorly overall with humans, completely missing the prevalence of \clusters and \smallTrends conclusions.
GPT-4o aligned more closely to responses from humans for line charts than other chart types, although it generated greater proportions of \clusters and \largeTrends takeaways.

\subsubsection{GPT-4o: Rank Charts}

GPT-4o emulated the bias towards line charts observed in humans via a strong preference for line charts. However, there was almost no variation in GPT-4o rankings; line charts were almost always ranked first ($91\%$ of rankings) by GPT-4o, dot plots ($91\%$), and heatmaps third ($99\%$). Using Durbin tests and post-hoc Conover testing with Holm correction, these differences were significant across all factors ($p<0.001$). 
GPT-4o demonstrated significant limitations in variability compared to human responses. 
Heatmaps were ranked first or second for only $0.01\%$ ($7$) of the GPT-4o rankings, while humans ranked them first or second for $24\%$ ($565$) of responses. 

Affordances across GPT-4o first-ranked charts partially aligned with human affordances.
The distribution of GPT-4o first-ranked charts was not uniform across factors, shown in \cref{fig:casestudyoverview}B ($\chi^2 = 89.15$, $df = 8$, $p < 0.001$). 
For both human and GPT-4o, first-ranked line charts best captured \smallTrends.
We could not extract GPT-4o affordances for heatmaps due to the low number of first-rankings compared to human responses.
GPT-4o first-ranked dotplots afforded \clusters, while human responses indicated that dotplots afforded \points.

\subsubsection{GPT-4o: Rank Takeaways}

GPT-4o generated overall poorly aligned relative takeaway rankings as compared to humans. 
The rankings elicited from this method differed greatly from rankings provided by humans, particularly for dot plots and line charts. As seen in~\cref{fig:casestudyoverview}C, GPT-4o demonstrated an undue preference for \largeTrends; human responses tended to rank \shape highest.
While \largeTrends were also ranked highly by humans, GPT-4o overrepresented the salience of \largeTrends with substantially higher proportions of first rankings for all chart types.
Likewise, GPT-4o failed to capture the relative affordance of \clusters for heatmaps and \smallTrends for line charts. While \points and \shape may be relatively afforded for dot plots (similar to human responses), the overall impact of \largeTrends dwarfs this observation.
Overall, affordances derived from across GPT-4o ranked takeaways aligned poorly with affordances from human rankings.

\subsubsection{GPT-4o: Rate Salience}

GPT-4o generated salience ratings that partially aligned with human responses. Post-hoc Tukey HSD testing revealed a significant main effect of chart type and factor for GPT-4o ratings ($p<0.001$).
Overall, \smallTrends takeaways received higher average salience ratings for humans and GPT-4o, see~\cref{fig:casestudyoverview}D. 
Takeaways paired with heatmaps received lower average ratings for both human and GPT-4o responses.

However, affordances by chart type differed slightly. 
We found two significant interactions between takeaway factor and chart type for GPT-4o responses; this interaction was not present for human responses. 
\smallTrends and \points takeaways were rated significantly more salient than \shape for heatmaps. 
Therefore, while saliency ratings for humans indicated no significant chart-specific affordances, GPT-4o seemed more sensitive to heatmaps. 
However, these affordances for heatmaps are not consistent with the overall findings from the human studies, which suggest that heatmaps afford \clusters.

\subsection{GPT-4o Evaluation}

Our results overall suggest that GPT-4o may not be a reliable proxy for human subjects when studying visualization affordances. However, analyzing output from each of the four study-based prompts resulted in useful insights into specifics of \textit{how} GPT-4o compared to humans. The prompt requesting free-response output highlighted distinct factual inaccuracies (predominantly with heatmaps) and prompts outlining the ranking tasks revealed strong model preferences for line charts and large trend conclusions. The prompt asking for salience ratings was the most promising for matching human responses.

\section{Discussion}
\label{discussion}

We evaluated multiple elicitation methods for identifying visualization affordances, with the goal of finding a more scalable alternative to free-response tasks. In this section, we synthesize takeaways for researchers selecting methods to study visualization affordances. 

\pheading{Methodology Trade-offs: Overall Salience vs. Specific Affordances.}
The structured methods we tested (ranking charts, ranking conclusions, and rating salience) each \textit{offered partial insight into visualization affordances but also introduced systematic limitations.}
These findings suggest that while quick methods may offer reasonable proxies, they are prone to biases that must be carefully considered. 
Line charts, for example, were consistently ranked highest across ranking tasks; this likely reflects genuine affordances for identifying trends but was also associated with familiarity, among other possible factors. 
If one were to evaluate these results in isolation, it would be tempting to conclude that line charts are universally the best chart type. 
However, this is not true (e.g., ~\cite{bertini2020shouldn}); Study 1 showed that different chart types elicit different types of conclusions.

Both ranking and rating methods revealed selective alignment with affordance patterns from the free-response study. 
Salience ratings aligned fairly closely with free-response frequencies. \smallTrends, for example, emerged as both common and highly salient. 
However, the lack of interaction effects in Study 4 indicates that participants rated factors similarly across chart types, limiting the ability to detect chart-specific affordances. 
Ranking tasks revealed more differences but introduced comparison effects.
For example, \shape conclusions were rarely generated in Study 1 but often ranked highly in Study 3, suggesting that comparison may have elevated their perceived salience.

Together, these studies suggest that ``what appears most salient'' and ``what takeaways come intuitively to people'' are not always equivalent. 
Research methodology will impact outcomes. 
Ranking and rating methods may highlight the most visually emphasized patterns in data, particularly when participants are presented with multiple options. In contrast, free-response tasks better capture what participants spontaneously derive from visualizations.

\pheading{Combining Methods for Studying Affordances.}
When reducing research overhead such as participant burden or effortful qualitative coding analysis, combining elicitation methods may offer a promising way forward.
\textit{Salience ratings and conclusion rankings both preserved some of the affordance signals found in free responses}, particularly for heatmaps and line charts. 
Taken together, these methods could triangulate patterns that approximate those found in open-ended tasks.
Combinations of ranking and rating methodologies can shed light on chart takeaways across different visualization designs. 
However, to fully understand patterns people see in data, human free-response remains the most complete method. 

\pheading{Specific Affordances for Chart Types.}
Based on the results from Studies 1-4, we can propose a set of affordances for the three chart types examined. 
\textit{Heatmaps most clearly afford \clusters and line charts afford \smallTrends.} Dot plots afford \shape conclusions under certain conditions, but this affordance was less consistent across studies; ranking procedures indicated that dot plots afforded \points. 
As such, \textit{we did not find converging evidence for specific affordances for dot plots}. Both \points and \shape could be potential affordances, depending on the elicitation method. 

We found that \largeTrends and \smallTrends conclusions were common in charts with increasing or decreasing trends, while \clusters and \shape were frequent when there was no net change. 
We found no interaction between chart type and data pattern. This suggests that certain data trends, which designers have little control over, may afford different takeaways, independent of visual encoding.

\pheading{Considerations for Evaluating LLMs as Human Proxies in Chart Interpretation.}
From our case study, we concluded that prompting GPT-4o to provide \textit{salience ratings for possible takeaways provided a comparable analysis of visualization affordances} but was less successful than human elicitation methods.
In addition, we found that our approach of testing GPT-4o with various prompts based on human study instructions resulted in a variety of insights that informed GPT-4o's capability to interpret charts in a human-like manner. Thus, we posit that future research related to LLM chart interpretation capabilities may benefit from similar explorations of LLM prompting based on human study methods. Evaluations of the model output can then involve a comparison of LLMs to humans, serving as a preliminary step to direct more refined prompt engineering efforts by researchers.

\section{Limitations and Future Work}

While our studies reveal useful patterns for evaluating visualization affordances, several limitations shape the interpretation of our findings and point to important directions for future work.

First, our studies focused on time series data, using only three chart types and fifteen datasets. 
While the chart types tested differed in encoding (position vs. color), we acknowledge they may not yield strongly contrasting affordances; future work could expand this approach to include richer chart types or include text elements to test whether stronger affordance signals emerge.
Future work should examine a broader range of visualization and data types. 

Additionally, participants viewed visualizations without context, which helped us isolate design effects but differed from real-world, task-driven interpretation.
Existing research has shown that user tasks can dictate takeaways from visualizations~\cite{malpica2023task}.
Our interpretation of affordances relied on factors derived from a preliminary study with time-series data. These factors may not fully reflect the range of interpretations users could generate for other task contexts or data types.
Future research could explore how elicitation methods vary for task-driven conditions (e.g., evacuation decisions in disasters) to assess further method-specific affordances.

Study 2, where participants compared multiple charts, mirrors real-world visualization recommendation tools~\cite{wongsuphasawat2017voyager}. 
Participants showed a preference for line charts, which were more familiar, highlighting the need for such tools to consider user biases in recommendation design.

While the primary focus of this work was on human interpretation, we also evaluated the use of LLMs as human proxies, using only a single prompt structure per method.
LLM output can be sensitive to changes in the prompt~\cite{liu2024lost}.
Future work should improve LLM responses through systematic prompt engineering using existing optimization tools and metrics~\cite{strobelt2022interactive,qin2021learning}.
Future work can also consider alternative tasks, such as asking LLMs to predict the takeaway of a specific person, given details about their characteristics (e.g., personality or literacy)~\cite{das2025leveraging}.

\section{Conclusion}

While structured methods like ranking and salience rating can approximate certain aspects of visualization affordances, they do not fully capture the richness or nuance of free-response from user studies.
Ranking tasks demonstrated how comparisons can shift perceptions of chart affordances; salience ratings failed to reflect chart-specific effects. 
These differences between methods reveal that designer choices, as well as researcher choices (i.e., how affordances are elicited), can shape what people take away from data and visualizations.





\acknowledgments{%
    The authors wish to thank Minsuk Chang, Yishu Ji, Will Wang, and Simone Laszuk for their support, 
    and the reviewers for their helpful feedback.
    This work was supported in part by the National Science Foundation Graduate Research Fellowship Grant No. DGE 2146752, as well as NSF awards IIS-2237585 and IIS-2311575. 
}

\bibliographystyle{abbrv-doi-hyperref-narrow}

\bibliography{references}

\begin{thebibliography}{10}
\renewcommand*{\sfdefault}{PTSansNarrow-TLF}

\bibitem{achiam2023gpt}
J.~Achiam, S.~Adler, S.~Agarwal, L.~Ahmad, I.~Akkaya, F.~L. Aleman, D.~Almeida, J.~Altenschmidt, S.~Altman, S.~Anadkat, et~al.
\newblock Gpt-4 technical report.
\newblock {\em arXiv preprint arXiv:2303.08774},  100 pages, 2023.

\bibitem{aickin1996adjusting}
M.~Aickin and H.~Gensler.
\newblock Adjusting for multiple testing when reporting research results: the bonferroni vs holm methods.
\newblock {\em American journal of public health}, 86(5):726--728, 1996.

\bibitem{ajani2021declutter}
K.~Ajani, E.~Lee, C.~Xiong, C.~N. Knaflic, W.~Kemper, and S.~Franconeri.
\newblock Declutter and focus: Empirically evaluating design guidelines for effective data communication.
\newblock {\em IEEE Transactions on Visualization and Computer Graphics}, 28(10):3351--3364, 2021.

\bibitem{amar2005low}
R.~Amar, J.~Eagan, and J.~Stasko.
\newblock Low-level components of analytic activity in information visualization.
\newblock In {\em Proceedings of the Proceedings of the 2005 IEEE Symposium on Information Visualization}, INFOVIS '05, p.~15. IEEE Computer Society, USA, 2005. \href{https://doi.org/10.1109/INFOVIS.2005.24}
{doi: \textsf{%
10\hspace{.1pt}\discretionary{.}{%
}{.}\hspace{.4pt}1109\discretionary{/}{%
}{/}INFOVIS\hspace{.1pt}\discretionary{.}{%
}{.}\hspace{.4pt}2005\hspace{.1pt}\discretionary{.}{%
}{.}\hspace{.4pt}24}}


\bibitem{argyle2023out}
L.~P. Argyle, E.~C. Busby, N.~Fulda, J.~R. Gubler, C.~Rytting, and D.~Wingate.
\newblock Out of one, many: Using language models to simulate human samples.
\newblock {\em Political Analysis}, 31(3):337--351, 2023.

\bibitem{battle2023insight}
L.~Battle and A.~Ottley.
\newblock What exactly is an insight? a literature review.
\newblock In {\em 2023 IEEE Visualization and Visual Analytics (VIS)}, pp. 91--95, 2023. \href{https://doi.org/10.1109/VIS54172.2023.00027}
{doi: \textsf{%
10\hspace{.1pt}\discretionary{.}{%
}{.}\hspace{.4pt}1109\discretionary{/}{%
}{/}VIS54172\hspace{.1pt}\discretionary{.}{%
}{.}\hspace{.4pt}2023\hspace{.1pt}\discretionary{.}{%
}{.}\hspace{.4pt}00027}}


\bibitem{bearfield2023does}
C.~X. Bearfield, C.~Stokes, A.~Lovett, and S.~Franconeri.
\newblock What does the chart say? grouping cues guide viewer comparisons and conclusions in bar charts.
\newblock {\em IEEE Transactions on Visualization and Computer Graphics}, 30(8):5097--5110, 2024. \href{https://doi.org/10.1109/TVCG.2023.3289292}
{doi: \textsf{%
10\hspace{.1pt}\discretionary{.}{%
}{.}\hspace{.4pt}1109\discretionary{/}{%
}{/}TVCG\hspace{.1pt}\discretionary{.}{%
}{.}\hspace{.4pt}2023\hspace{.1pt}\discretionary{.}{%
}{.}\hspace{.4pt}3289292}}


\bibitem{bendeck2024empirical}
A.~Bendeck and J.~Stasko.
\newblock An empirical evaluation of the gpt-4 multimodal language model on visualization literacy tasks.
\newblock {\em IEEE VIS}, pp. 1--11, 2024.
\newblock to appear. \href{https://doi.org/10.1109/TVCG.2024.3456155}
{doi: \textsf{%
10\hspace{.1pt}\discretionary{.}{%
}{.}\hspace{.4pt}1109\discretionary{/}{%
}{/}TVCG\hspace{.1pt}\discretionary{.}{%
}{.}\hspace{.4pt}2024\hspace{.1pt}\discretionary{.}{%
}{.}\hspace{.4pt}3456155}}


\bibitem{bertin1983semiology}
J.~Bertin.
\newblock Semiology of graphics; diagrams networks maps.
\newblock Technical report, 1983.

\bibitem{bertini2020shouldn}
E.~Bertini, M.~Correll, and S.~Franconeri.
\newblock Why shouldn’t all charts be scatter plots? beyond precision-driven visualizations.
\newblock In {\em 2020 IEEE Visualization Conference (VIS)}, pp. 206--210. IEEE Computer Society, Los Alamitos, CA, USA, oct 2020. \href{https://doi.org/10.1109/VIS47514.2020.00048}
{doi: \textsf{%
10\hspace{.1pt}\discretionary{.}{%
}{.}\hspace{.4pt}1109\discretionary{/}{%
}{/}VIS47514\hspace{.1pt}\discretionary{.}{%
}{.}\hspace{.4pt}2020\hspace{.1pt}\discretionary{.}{%
}{.}\hspace{.4pt}00048}}


\bibitem{borkin2013makes}
M.~A. Borkin, A.~A. Vo, Z.~Bylinskii, P.~Isola, S.~Sunkavalli, A.~Oliva, and H.~Pfister.
\newblock What makes a visualization memorable?
\newblock {\em IEEE Transactions on Visualization and Computer Graphics}, 19(12):2306--2315, 2013.

\bibitem{boy2015suggested}
J.~Boy, L.~Eveillard, F.~Detienne, and J.-D. Fekete.
\newblock Suggested interactivity: Seeking perceived affordances for information visualization.
\newblock {\em IEEE Transactions on Visualization and Computer Graphics}, 22(1):639--648, 2015.

\bibitem{brehmer2013multi}
M.~Brehmer and T.~Munzner.
\newblock A multi-level typology of abstract visualization tasks.
\newblock {\em IEEE Transactions on Visualization and Computer Graphics}, 19(12):2376--2385, 2013.

\bibitem{burns2020evaluate}
A.~Burns, C.~Xiong, S.~Franconeri, A.~Cairo, and N.~Mahyar.
\newblock How to evaluate data visualizations across different levels of understanding.
\newblock In {\em 2020 IEEE Workshop on Evaluation and Beyond - Methodological Approaches to Visualization (BELIV)}, pp. 19--28. IEEE Computer Society, Los Alamitos, CA, USA, oct 2020. \href{https://doi.org/10.1109/BELIV51497.2020.00010}
{doi: \textsf{%
10\hspace{.1pt}\discretionary{.}{%
}{.}\hspace{.4pt}1109\discretionary{/}{%
}{/}BELIV51497\hspace{.1pt}\discretionary{.}{%
}{.}\hspace{.4pt}2020\hspace{.1pt}\discretionary{.}{%
}{.}\hspace{.4pt}00010}}


\bibitem{cabouat2024previs}
A.-F. Cabouat, T.~He, P.~Isenberg, and T.~Isenberg.
\newblock Previs: Perceived readability evaluation for visualizations.
\newblock {\em IEEE Transactions on Visualization and Computer Graphics}, 2024.

\bibitem{chen2024viseval}
N.~Chen, Y.~Zhang, J.~Xu, K.~Ren, and Y.~Yang.
\newblock Viseval: A benchmark for data visualization in the era of large language models.
\newblock {\em IEEE VIS},  19 pages, 2024.
\newblock to appear.

\bibitem{cleveland1984graphical}
W.~S. Cleveland and R.~McGill.
\newblock Graphical perception: Theory, experimentation, and application to the development of graphical methods.
\newblock {\em Journal of the American Statistical Association}, 79(387):531--554, 1984.

\bibitem{crouser2012affordance}
R.~J. Crouser and R.~Chang.
\newblock An affordance-based framework for human computation and human-computer collaboration.
\newblock {\em IEEE Transactions on Visualization and Computer Graphics}, 18(12):2859--2868, 2012.

\bibitem{cui2024promises}
Y.~Cui, L.~W. Ge, Y.~Ding, L.~Harrison, F.~Yang, and M.~Kay.
\newblock Promises and pitfalls: Using large language models to generate visualization items.
\newblock {\em IEEE Transactions on Visualization and Computer Graphics}, pp. 1--11, 2024. \href{https://doi.org/10.1109/TVCG.2024.3456309}
{doi: \textsf{%
10\hspace{.1pt}\discretionary{.}{%
}{.}\hspace{.4pt}1109\discretionary{/}{%
}{/}TVCG\hspace{.1pt}\discretionary{.}{%
}{.}\hspace{.4pt}2024\hspace{.1pt}\discretionary{.}{%
}{.}\hspace{.4pt}3456309}}


\bibitem{das2025leveraging}
A.~K. Das, K.~Mueller, and C.~Xiong.
\newblock Leveraging large language models for personalized public messaging.
\newblock In {\em Proceedings of the 2025 CHI Conference on Human Factors in Computing Systems Extended Abstracts (CHI LBW)}. ACM, Honolulu, HI, USA, 2025.
\newblock Late-Breaking Work.

\bibitem{dillion2023can}
D.~Dillion, N.~Tandon, Y.~Gu, and K.~Gray.
\newblock Can ai language models replace human participants?
\newblock {\em Trends in Cognitive Sciences}, 27(7):597--600, 2023.

\bibitem{elhamdadi2023vistrust}
H.~Elhamdadi, A.~Stefkovics, J.~Beyer, E.~Moerth, H.~Pfister, C.~X. Bearfield, and C.~Nobre.
\newblock Vistrust: a multidimensional framework and empirical study of trust in data visualizations.
\newblock {\em IEEE Transactions on Visualization and Computer Graphics}, 30(1):348--358, 2023.

\bibitem{faul2007g}
F.~Faul, E.~Erdfelder, A.-G. Lang, and A.~Buchner.
\newblock G* power 3: A flexible statistical power analysis program for the social, behavioral, and biomedical sciences.
\newblock {\em Behavior research methods}, 39(2):175--191, 2007.

\bibitem{forsell2010heuristic}
C.~Forsell and J.~Johansson.
\newblock An heuristic set for evaluation in information visualization.
\newblock In {\em Proceedings of the International Conference on Advanced Visual Interfaces}, pp. 199--206, 2010.

\bibitem{fygenson2023arrangement}
R.~Fygenson, S.~Franconeri, and E.~Bertini.
\newblock The arrangement of marks impacts afforded messages: Ordering, partitioning, spacing, and coloring in bar charts.
\newblock {\em IEEE Transactions on Visualization and Computer Graphics}, 30(01):1008--1018, jan 2024. \href{https://doi.org/10.1109/TVCG.2023.3326590}
{doi: \textsf{%
10\hspace{.1pt}\discretionary{.}{%
}{.}\hspace{.4pt}1109\discretionary{/}{%
}{/}TVCG\hspace{.1pt}\discretionary{.}{%
}{.}\hspace{.4pt}2023\hspace{.1pt}\discretionary{.}{%
}{.}\hspace{.4pt}3326590}}


\bibitem{gilardi2023chatgpt}
F.~Gilardi, M.~Alizadeh, and M.~Kubli.
\newblock Chatgpt outperforms crowd workers for text-annotation tasks.
\newblock {\em Proceedings of the National Academy of Sciences}, 120(30):e2305016120, 2023.

\bibitem{gleicher2011visual}
M.~Gleicher, D.~Albers, R.~Walker, I.~Jusufi, C.~D. Hansen, and J.~C. Roberts.
\newblock Visual comparison for information visualization.
\newblock {\em Information Visualization}, 10(4):289–309,  21 pages, oct 2011. \href{https://doi.org/10.1177/1473871611416549}
{doi: \textsf{%
10\hspace{.1pt}\discretionary{.}{%
}{.}\hspace{.4pt}1177\discretionary{/}{%
}{/}1473871611416549}}


\bibitem{goethals2024one}
S.~Goethals and L.~Rhue.
\newblock One world, one opinion? the superstar effect in llm responses.
\newblock {\em arXiv preprint arXiv:2412.10281}, 2024.

\bibitem{gutierrez2019benefits}
F.~Guti{\'e}rrez, X.~Ochoa, K.~Seipp, T.~Broos, and K.~Verbert.
\newblock Benefits and trade-offs of different model representations in decision support systems for non-expert users.
\newblock In {\em IFIP TC13 International Conference on Human-Computer Interaction}, pp. 576--597. Springer, 2019.

\bibitem{hamalainen2023evaluating}
P.~H{\"a}m{\"a}l{\"a}inen, M.~Tavast, and A.~Kunnari.
\newblock Evaluating large language models in generating synthetic hci research data: a case study.
\newblock In {\em Proceedings of the 2023 CHI Conference on Human Factors in Computing Systems}, pp. 1--19, 2023.

\bibitem{harding2024ai}
J.~Harding, W.~D’Alessandro, N.~Laskowski, and R.~Long.
\newblock Ai language models cannot replace human research participants.
\newblock {\em Ai \& Society}, 39(5):2603--2605, 2024.

\bibitem{harrison2014ranking}
L.~Harrison, F.~Yang, S.~Franconeri, and R.~Chang.
\newblock {Ranking Visualizations of Correlation Using Weber's Law}.
\newblock {\em IEEE Transactions on Visualization and Computer Graphics}, 20(12):1943--1952, 2014. \href{https://doi.org/10.1109/tvcg.2014.2346979}
{doi: \textsf{%
10\hspace{.1pt}\discretionary{.}{%
}{.}\hspace{.4pt}1109\discretionary{/}{%
}{/}tvcg\hspace{.1pt}\discretionary{.}{%
}{.}\hspace{.4pt}2014\hspace{.1pt}\discretionary{.}{%
}{.}\hspace{.4pt}2346979}}


\bibitem{he2022beauvis}
T.~He, P.~Isenberg, R.~Dachselt, and T.~Isenberg.
\newblock Beauvis: A validated scale for measuring the aesthetic pleasure of visual representations.
\newblock {\em IEEE Transactions on Visualization and Computer Graphics}, 2022.

\bibitem{hearst2016evaluating}
M.~A. Hearst, P.~Laskowski, and L.~Silva.
\newblock Evaluating information visualization via the interplay of heuristic evaluation and question-based scoring.
\newblock In {\em Proceedings of the 2016 CHI Conference on Human Factors in Computing Systems}, pp. 5028--5033, 2016.

\bibitem{heider2011local}
P.~Heider, A.~Pierre-Pierre, R.~Li, and C.~Grimm.
\newblock Local shape descriptors, a survey and evaluation.
\newblock In {\em Proceedings of the 4th Eurographics Conference on 3D Object Retrieval}, 3DOR '11,  8 pages, p. 49–56. Eurographics Association, Goslar, DEU, 2011.

\bibitem{holder2022dispersion}
E.~Holder and C.~Xiong.
\newblock Dispersion vs disparity: Hiding variability can encourage stereotyping when visualizing social outcomes.
\newblock {\em IEEE Transactions on Visualization and Computer Graphics}, 29(1):624--634, 2022.

\bibitem{hong2025llms}
J.~Hong, C.~Seto, A.~Fan, and R.~Maciejewski.
\newblock Do llms have visualization literacy? an evaluation on modified visualizations to test generalization in data interpretation.
\newblock {\em IEEE Transactions on Visualization and Computer Graphics}, 2025.

\bibitem{horton2023large}
J.~J. Horton.
\newblock Large language models as simulated economic agents: What can we learn from homo silicus?
\newblock Technical report, National Bureau of Economic Research, 2023.

\bibitem{kay2016ish}
M.~Kay, T.~Kola, J.~R. Hullman, and S.~A. Munson.
\newblock When (ish) is my bus? user-centered visualizations of uncertainty in everyday, mobile predictive systems.
\newblock In {\em Proceedings of the 2016 CHI Conference on Human Factors in Computing Systems}, CHI '16,  12 pages, p. 5092–5103. Association for Computing Machinery, New York, NY, USA, 2016. \href{https://doi.org/10.1145/2858036.2858558}
{doi: \textsf{%
10\hspace{.1pt}\discretionary{.}{%
}{.}\hspace{.4pt}1145\discretionary{/}{%
}{/}2858036\hspace{.1pt}\discretionary{.}{%
}{.}\hspace{.4pt}2858558}}


\bibitem{lin2025proxy}
K.~Lin, C.~Stokes, and C.~X. Bearfield.
\newblock Llms are not reliable human proxies to study affordances in data visualizations.
\newblock In {\em CHI 2025 Workshop on Human-centered Evaluation and Auditing of Language Models},  8 pages. Association for Computing Machinery, 2025.

\bibitem{liu2024lost}
N.~F. Liu, K.~Lin, J.~Hewitt, A.~Paranjape, M.~Bevilacqua, F.~Petroni, and P.~Liang.
\newblock Lost in the middle: How language models use long contexts.
\newblock {\em Transactions of the Association for Computational Linguistics}, 12:157--173, 2024.

\bibitem{liu2014survey}
S.~Liu, W.~Cui, Y.~Wu, and M.~Liu.
\newblock A survey on information visualization: recent advances and challenges.
\newblock {\em The Visual Computer}, 30:1373--1393, 2014.

\bibitem{liu2023fill}
Z.~Liu, C.~Chen, J.~Wang, X.~Che, Y.~Huang, J.~Hu, and Q.~Wang.
\newblock Fill in the blank: Context-aware automated text input generation for mobile gui testing.
\newblock In {\em Proceedings of the 45th International Conference on Software Engineering}, ICSE '23,  13 pages, p. 1355–1367. IEEE Press, Melbourne, Victoria, Australia, 2023. \href{https://doi.org/10.1109/ICSE48619.2023.00119}
{doi: \textsf{%
10\hspace{.1pt}\discretionary{.}{%
}{.}\hspace{.4pt}1109\discretionary{/}{%
}{/}ICSE48619\hspace{.1pt}\discretionary{.}{%
}{.}\hspace{.4pt}2023\hspace{.1pt}\discretionary{.}{%
}{.}\hspace{.4pt}00119}}


\bibitem{malpica2023task}
S.~Malpica, D.~Martin, A.~Serrano, D.~Gutierrez, and B.~Masia.
\newblock Task-dependent visual behavior in immersive environments: A comparative study of free exploration, memory and visual search.
\newblock {\em IEEE Transactions on Visualization and Computer Graphics}, 29(11):4417--4425, 2023. \href{https://doi.org/10.1109/TVCG.2023.3320259}
{doi: \textsf{%
10\hspace{.1pt}\discretionary{.}{%
}{.}\hspace{.4pt}1109\discretionary{/}{%
}{/}TVCG\hspace{.1pt}\discretionary{.}{%
}{.}\hspace{.4pt}2023\hspace{.1pt}\discretionary{.}{%
}{.}\hspace{.4pt}3320259}}


\bibitem{matzen2023numerical}
L.~E. Matzen, B.~C. Howell, M.~C.~S. Trumbo, and K.~M. Divis.
\newblock Numerical and visual representations of uncertainty lead to different patterns of decision making.
\newblock {\em IEEE Computer Graphics and Applications}, 43(5):72--82, 2023. \href{https://doi.org/10.1109/MCG.2023.3299875}
{doi: \textsf{%
10\hspace{.1pt}\discretionary{.}{%
}{.}\hspace{.4pt}1109\discretionary{/}{%
}{/}MCG\hspace{.1pt}\discretionary{.}{%
}{.}\hspace{.4pt}2023\hspace{.1pt}\discretionary{.}{%
}{.}\hspace{.4pt}3299875}}


\bibitem{munzner2014visualization}
T.~Munzner.
\newblock {\em Visualization analysis and design}.
\newblock CRC press, 2014.

\bibitem{navon1977forest}
D.~Navon.
\newblock Forest before trees: The precedence of global features in visual perception.
\newblock {\em Cognitive psychology}, 9(3):353--383, 1977.

\bibitem{padilla2018decision}
L.~M. Padilla, S.~H. Creem-Regehr, M.~Hegarty, and J.~K. Stefanucci.
\newblock Decision making with visualizations: a cognitive framework across disciplines.
\newblock {\em Cognitive research: principles and implications}, 3(1):29, 2018.

\bibitem{padilla2020powerful}
L.~M. Padilla, S.~H. Creem-Regehr, and W.~Thompson.
\newblock The powerful influence of marks: Visual and knowledge-driven processing in hurricane track displays.
\newblock {\em Journal of experimental psychology: applied}, 26(1):1, 2020.

\bibitem{palan2018prolific}
S.~Palan and C.~Schitter.
\newblock Prolific. ac—a subject pool for online experiments.
\newblock {\em Journal of Behavioral and Experimental Finance}, 17:22--27, 2018.

\bibitem{pandey2023you}
S.~Pandey, O.~G. McKinley, R.~J. Crouser, and A.~Ottley.
\newblock Do you trust what you see? toward a multidimensional measure of trust in visualization.
\newblock In {\em 2023 IEEE Visualization and Visual Analytics (VIS)}, pp. 26--30. IEEE, 2023.

\bibitem{plaisant2008insight}
C.~Plaisant, J.-D. Fekete, and G.~Grinstein.
\newblock Promoting insight-based evaluation of visualizations: From contest to benchmark repository.
\newblock {\em IEEE Transactions on Visualization and Computer Graphics}, 14(1):120--134, 2008. \href{https://doi.org/10.1109/TVCG.2007.70412}
{doi: \textsf{%
10\hspace{.1pt}\discretionary{.}{%
}{.}\hspace{.4pt}1109\discretionary{/}{%
}{/}TVCG\hspace{.1pt}\discretionary{.}{%
}{.}\hspace{.4pt}2007\hspace{.1pt}\discretionary{.}{%
}{.}\hspace{.4pt}70412}}


\bibitem{qin2021learning}
G.~Qin and J.~Eisner.
\newblock Learning how to ask: Querying {LM}s with mixtures of soft prompts.
\newblock In {\em Proceedings of the 2021 Conference of the North American Chapter of the Association for Computational Linguistics: Human Language Technologies}, pp. 5203--5212. Association for Computational Linguistics, 2021. \href{https://doi.org/10.18653/v1/2021.naacl-main.410}
{doi: \textsf{%
10\hspace{.1pt}\discretionary{.}{%
}{.}\hspace{.4pt}18653\discretionary{/}{%
}{/}v1\discretionary{/}{%
}{/}2021\hspace{.1pt}\discretionary{.}{%
}{.}\hspace{.4pt}naacl\discretionary{%
}{-}{-}main\hspace{.1pt}\discretionary{.}{%
}{.}\hspace{.4pt}410}}


\bibitem{rensink2010perception}
R.~A. Rensink and G.~Baldridge.
\newblock The perception of correlation in scatterplots.
\newblock In {\em Computer graphics forum}, vol.~29, pp. 1203--1210. Wiley Online Library, 2010.

\bibitem{revelle2015package}
W.~Revelle and M.~W. Revelle.
\newblock Package ‘psych’.
\newblock {\em The comprehensive R archive network}, 337:338, 2015.

\bibitem{saraiya2005insight}
P.~Saraiya, C.~North, and K.~Duca.
\newblock An insight-based methodology for evaluating bioinformatics visualizations.
\newblock {\em IEEE Transactions on Visualization and Computer Graphics}, 11(4):443--456, 2005. \href{https://doi.org/10.1109/TVCG.2005.53}
{doi: \textsf{%
10\hspace{.1pt}\discretionary{.}{%
}{.}\hspace{.4pt}1109\discretionary{/}{%
}{/}TVCG\hspace{.1pt}\discretionary{.}{%
}{.}\hspace{.4pt}2005\hspace{.1pt}\discretionary{.}{%
}{.}\hspace{.4pt}53}}


\bibitem{schapira2001frequency}
M.~M. Schapira, A.~B. Nattinger, and C.~A. McHorney.
\newblock Frequency or probability? a qualitative study of risk communication formats used in health care.
\newblock {\em Medical Decision Making}, 21(6):459--467, 2001.

\bibitem{schloss2024color}
K.~B. Schloss.
\newblock Color semantics in human cognition.
\newblock {\em Current Directions in Psychological Science}, 33(1):58--67, 2024.

\bibitem{schulz2013design}
H.-J. Schulz, T.~Nocke, M.~Heitzler, and H.~Schumann.
\newblock A design space of visualization tasks.
\newblock {\em IEEE Transactions on Visualization and Computer Graphics}, 19(12):2366--2375, 2013.

\bibitem{setlur2015linguistic}
V.~Setlur and M.~C. Stone.
\newblock A linguistic approach to categorical color assignment for data visualization.
\newblock {\em IEEE Transactions on Visualization and Computer Graphics}, 22(1):698--707, 2015.

\bibitem{shneiderman2003eyes}
B.~Shneiderman.
\newblock The eyes have it: A task by data type taxonomy for information visualizations.
\newblock In {\em The Craft of Information Visualization}, pp. 364--371. Elsevier, 2003.

\bibitem{slomska2021different}
K.~S{\l}omska-Przech and I.~M. Go{\l}{\k{e}}biowska.
\newblock Do different map types support map reading equally? comparing choropleth, graduated symbols, and isoline maps for map use tasks.
\newblock {\em ISPRS International Journal of Geo-Information}, 10(2):69, 2021.

\bibitem{snow2013qualtrics}
J.~Snow and M.~Mann.
\newblock Qualtrics survey software: handbook for research professionals.
\newblock {\em Provo, UT: Qualtrics Labs},  209 pages, 2013.

\bibitem{south2022effective}
L.~South, D.~Saffo, O.~Vitek, C.~Dunne, and M.~A. Borkin.
\newblock Effective use of likert scales in visualization evaluations: A systematic review.
\newblock In {\em Computer Graphics Forum}, vol.~41, pp. 43--55. Wiley Online Library, 2022.

\bibitem{southwell2022defining}
B.~G. Southwell, J.~S.~B. Brennen, R.~Paquin, V.~Boudewyns, and J.~Zeng.
\newblock Defining and measuring scientific misinformation.
\newblock {\em The ANNALS of the American Academy of Political and Social Science}, 700(1):98--111, 2022.

\bibitem{stokes2024delays}
C.~Stokes, C.~Sanker, B.~Cogley, and V.~Setlur.
\newblock {From Delays to Densities: Exploring Data Uncertainty through Speech, Text, and Visualization}.
\newblock In {\em Computer Graphics Forum}, vol.~43, p. e15100. Wiley Online Library, 2024. \href{https://doi.org/10.1111/cgf.15100}
{doi: \textsf{%
10\hspace{.1pt}\discretionary{.}{%
}{.}\hspace{.4pt}1111\discretionary{/}{%
}{/}cgf\hspace{.1pt}\discretionary{.}{%
}{.}\hspace{.4pt}15100}}


\bibitem{stokes2022balance}
C.~Stokes, V.~Setlur, B.~Cogley, A.~Satyanarayan, and M.~A. Hearst.
\newblock Striking a balance: reader takeaways and preferences when integrating text and charts.
\newblock {\em IEEE Transactions on Visualization and Computer Graphics}, 29(1):1233--1243, 2022.

\bibitem{strobelt2022interactive}
H.~Strobelt, A.~Webson, V.~Sanh, B.~Hoover, J.~Beyer, H.~Pfister, and A.~M. Rush.
\newblock Interactive and visual prompt engineering for ad-hoc task adaptation with large language models.
\newblock {\em IEEE Transactions on Visualization and Computer Graphics}, 29(1):1146--1156, 2022.

\bibitem{szafir2016four}
D.~A. Szafir, S.~Haroz, M.~Gleicher, and S.~Franconeri.
\newblock Four types of ensemble coding in data visualizations.
\newblock {\em Journal of Vision}, 16(5):11--11, 2016.

\bibitem{tory2004rethinking}
M.~Tory and T.~Moller.
\newblock Rethinking visualization: A high-level taxonomy.
\newblock In {\em IEEE Symposium on Information Visualization}, pp. 151--158. IEEE, Austin, TX, USA, 2004.

\bibitem{treisman1982perceptual}
A.~Treisman.
\newblock Perceptual grouping and attention in visual search for features and for objects.
\newblock {\em Journal of experimental psychology: human perception and performance}, 8(2):194, 1982.

\bibitem{tseng2024shape}
C.~Tseng, A.~Z. Wang, G.~J. Quadri, and D.~A. Szafir.
\newblock Shape it up: An empirically grounded approach for designing shape palettes.
\newblock {\em IEEE VIS},  11 pages, 2024.
\newblock to appear.

\bibitem{tukey1977exploratory}
J.~W. Tukey et~al.
\newblock {\em Exploratory data analysis}, vol.~2.
\newblock Springer, 1977.

\bibitem{wang2024aligned}
H.~W. Wang, J.~Hoffswell, S.~M.~T. Thane, V.~S. Bursztyn, and C.~X. Bearfield.
\newblock How aligned are human chart takeaways and llm predictions? a case study on bar charts with varying layouts.
\newblock {\em IEEE VIS},  11 pages, 2024.
\newblock to appear.

\bibitem{wongsuphasawat2017voyager}
K.~Wongsuphasawat, Z.~Qu, D.~Moritz, R.~Chang, F.~Ouk, A.~Anand, J.~Mackinlay, B.~Howe, and J.~Heer.
\newblock Voyager 2: Augmenting visual analysis with partial view specifications.
\newblock In {\em Proceedings of the 2017 chi conference on human factors in computing systems}, pp. 2648--2659, 2017.

\bibitem{xiong2021visual}
C.~Xiong, V.~Setlur, B.~Bach, E.~Koh, K.~Lin, and S.~Franconeri.
\newblock Visual arrangements of bar charts influence comparisons in viewer takeaways.
\newblock {\em IEEE Transactions on Visualization and Computer Graphics}, 28(1):955--965, 2021.

\bibitem{xu2024exploring}
Z.~Xu and E.~Wall.
\newblock Exploring the capability of llms in performing low-level visual analytic tasks on svg data visualizations.
\newblock {\em IEEE VIS},  5 pages, 2024.

\bibitem{yang2023swaying}
F.~Yang, M.~Cai, C.~Mortenson, H.~Fakhari, A.~D. Lokmanoglu, J.~Hullman, S.~Franconeri, N.~Diakopoulos, E.~C. Nisbet, and M.~Kay.
\newblock Swaying the public? impacts of election forecast visualizations on emotion, trust, and intention in the 2022 u.s. midterms.
\newblock {\em IEEE Transactions on Visualization and Computer Graphics}, 30(01):23--33, jan 2024. \href{https://doi.org/10.1109/TVCG.2023.3327356}
{doi: \textsf{%
10\hspace{.1pt}\discretionary{.}{%
}{.}\hspace{.4pt}1109\discretionary{/}{%
}{/}TVCG\hspace{.1pt}\discretionary{.}{%
}{.}\hspace{.4pt}2023\hspace{.1pt}\discretionary{.}{%
}{.}\hspace{.4pt}3327356}}


\bibitem{zacks1999bars}
J.~Zacks and B.~Tversky.
\newblock Bars and lines: A study of graphic communication.
\newblock {\em Memory \& Cognition}, 27(6):1073--1079, 1999.

\bibitem{zha2023tablegpt}
L.~Zha, J.~Zhou, L.~Li, R.~Wang, Q.~Huang, S.~Yang, J.~Yuan, C.~Su, X.~Li, A.~Su, et~al.
\newblock Tablegpt: Towards unifying tables, nature language and commands into one gpt.
\newblock {\em arXiv preprint arXiv:2307.08674},  13 pages, 2023.

\bibitem{zhao2024leva}
Y.~Zhao, Y.~Zhang, Y.~Zhang, X.~Zhao, J.~Wang, Z.~Shao, C.~Turkay, and S.~Chen.
\newblock Leva: Using large language models to enhance visual analytics.
\newblock {\em IEEE Transactions on Visualization and Computer Graphics}, pp. 1--17, 2024. \href{https://doi.org/10.1109/TVCG.2024.3368060}
{doi: \textsf{%
10\hspace{.1pt}\discretionary{.}{%
}{.}\hspace{.4pt}1109\discretionary{/}{%
}{/}TVCG\hspace{.1pt}\discretionary{.}{%
}{.}\hspace{.4pt}2024\hspace{.1pt}\discretionary{.}{%
}{.}\hspace{.4pt}3368060}}


\bibitem{zohrevandi2022design}
E.~Zohrevandi, C.~A. Westin, J.~Lundberg, and A.~Ynnerman.
\newblock Design and evaluation study of visual analytics decision support tools in air traffic control.
\newblock In {\em Computer Graphics Forum}, vol.~41, pp. 230--242. Wiley Online Library, 2022.

\end{thebibliography}
\end{document}